%% file: sample-sigconf.tex
\documentclass[sigconf,screen]{acmart}

\usepackage{color}
\usepackage{xcolor}
\usepackage{hyperref}
\hypersetup{hidelinks=true} 
\usepackage{booktabs} 
\usepackage{enumitem}
\usepackage{amsmath,scalerel}
\usepackage{xspace}
\usepackage{graphicx}
\usepackage{balance} 
\usepackage[ruled,lined,linesnumbered,vlined,algo2e]{algorithm2e}   
\usepackage{listings}
\usepackage{multirow}

\newcommand{\cf}{\hbox{\emph{cf.}}\xspace}

\newcommand{\etal}{\hbox{\emph{et al.}}\xspace}
\newcommand{\eg}{\hbox{\emph{e.g.}}\xspace}
\newcommand{\ie}{\hbox{\emph{i.e.}}\xspace}

\newcommand{\wrt}{\hbox{\emph{w.r.t.}}\xspace}
\newcommand{\etc}{\hbox{\emph{etc}}\xspace}

\definecolor{light-gray}{gray}{0.8}

\newcommand{\tool}{{\textsf{\small{APEChecker}}}\xspace}

\makeatletter
\newcommand*\bigcdot{\mathpalette\bigcdot@{1}}
\newcommand*\bigcdot@[2]{\mathbin{\vcenter{\hbox{\scalebox{#2}{$\m@th#1\bullet$}}}}}
\makeatother

\copyrightyear{2018}
\acmYear{2018}
\setcopyright{acmcopyright}
\acmConference[ASE '18]{Proceedings of the 2018 33rd ACM/IEEE International Conference on Automated Software Engineering}{September 3--7, 2018}{Montpellier, France}
\acmBooktitle{Proceedings of the 2018 33rd ACM/IEEE International Conference on Automated Software Engineering (ASE '18), September 3--7, 2018, Montpellier, France}
\acmPrice{15.00}
\acmDOI{10.1145/3238147.3238170}
\acmISBN{978-1-4503-5937-5/18/09}

\begin{document}

\title{Efficiently Manifesting Asynchronous Programming Errors in Android Apps}

\author{Lingling Fan}
\affiliation{%
  \institution{East China Normal University, China}
  \streetaddress{}
  \city{}
  \country{}
  \postcode{}
}

\author{Ting Su}
\authornote{Ting Su is the corresponding author of this paper. \\Emails: ecnujanefan@gmail.com,  tsuletgo@gmail.com, ecnuchensen@gmail.com, gzmeng@ntu.edu.sg, yangliu@ntu.edu.sg, lihua.xu@nyu.edu, ggpu@sei.ecnu.edu.cn}
\affiliation{%
  \institution{East China Normal University, China \\ Nanyang Technological University, Singapore}
  \streetaddress{}
  \city{}
  \country{}
  \postcode{}
}

\author{Sen Chen}
\affiliation{%
  \institution{East China Normal University, China}
  \streetaddress{}
  \city{}
  \country{}
  }

\author{Guozhu Meng}
\affiliation{%
  \institution{Chinese Academy of Sciences, China \\ Nanyang Technological University, Singapore}
  \city{}
  \country{}
}

\author{Yang Liu}
\affiliation{%
 \institution{Nanyang Technological University, Singapore}
 \streetaddress{}
 \city{}
 \state{}
 \country{}
 }

\author{Lihua Xu}
\affiliation{%
  \institution{New York University Shanghai, China}
  \streetaddress{}
  \city{}
  \state{}
  \country{}
}

\author{Geguang Pu}
\affiliation{%
  \institution{East China Normal University, China}
  \streetaddress{}
  \city{}
  \state{}
  \country{}
  \postcode{}}

\renewcommand{\shortauthors}{L.~Fan, T.~Su, S.~Chen, G.~Meng, Y.~Liu, L.~Xu, and G.~Pu}

\begin{abstract}
Android, the \#1 mobile app framework, enforces the \emph{single-GUI-thread} model,
in which a single UI thread manages GUI rendering and event dispatching.
Due to this model, it is vital to avoid blocking the UI thread for responsiveness. One common practice is to offload long-running tasks into async threads. To achieve this, Android provides various async programming constructs, and leaves developers themselves to obey the rules implied by the model. However, as our study reveals, more than 25\% apps violate these rules and introduce hard-to-detect, fail-stop errors, which we term as aysnc programming errors (APEs). To this end, this paper introduces \tool, a technique to automatically and efficiently manifest APEs. The key idea is to characterize APEs as specific fault patterns, and synergistically combine static analysis and dynamic UI exploration to detect and verify such errors.
Among the 40 real-world Android apps, \tool unveils and processes 61 APEs, of which 51 are confirmed (83.6\% hit rate). Specifically, \tool detects 3X more APEs than the state-of-art testing tools (Monkey, Sapienz and Stoat), and reduces testing time from half an hour to a few minutes. On a specific type of APEs, \tool confirms 5X more errors than the data race detection tool, EventRacer, with very few false alarms.
\end{abstract}

%
%


\maketitle

\input{introduction.tex}
\input{overview.tex}

\input{approach.tex}
\input{evaluation.tex}
\input{relwork.tex}
\input{conclusion.tex}

\vspace{-3mm}
\begin{acks}
We appreciate the reviewers' constructive feedback. This work is partially supported by NSFC Grant 61502170, NTU Research Grant NGF-2017-03-033 and NRF Grant CRDCG2017-S04.
Lingling Fan is partly supported by ECNU Project of Funding Overseas Short-term Studies,
Ting Su is partially supported by NSFC Grant 61572197 and 61632005, and Geguang Pu is partially supported by 
MOST NKTSP Project 2015BAG19B02 and STCSM Project 16DZ1100600.

\end{acks}

\clearpage
\bibliographystyle{ACM-Reference-Format}
\balance
\bibliography{sample-sigconf}

\end{document}

%% file: introduction.tex
\vspace{-2mm}
\section{Introduction}
\label{sec:intro}

Most modern GUI frameworks such as \textsf{\small{Swing}}~\cite{Swing}, \textsf{\small{SWT}}~\cite{swt}, \textsf{\small{Android}}~\cite{Android}, \textsf{\small{Qt}}~\cite{Qt}, \textsf{\small{WxErlang}}~\cite{wxerlang}, and \textsf{\small{MacOS Cocoa}}~\cite{macos} enforce the \emph{single-GUI-thread model}, in which one single UI thread instantiates GUI components and dispatches events. Specifically, UI thread fetches system or user events off an event queue, and dispatches them either to a responsible app component's handler or to a UI widget's event handler. These handlers run on the UI thread and exclusively update GUIs if necessary.

Due to this single-GUI-thread model, it is vital to avoid blocking the UI thread. Therefore, most GUI frameworks recommend to offload intensive tasks (\eg, network access, database queries) to async threads (\ie, background threads).
Take Android development framework (ADF) as an example, it provides many async programming constructs (\eg, \textsf{\small{AysncTask}}, \textsf{\small{Thread}}, \textsf{\small{AsyncTaskLoader}}, \textsf{\small{IntentService}}) to achieve this goal.

Like other frameworks, ADF leaves developers themselves to properly handle the interactions between these async threads and the UI thread --- obey the rules implied by the single-UI-thread model.
However, our investigation on 930 apps that use async constructs shows, more than 25\% apps violate these rules, and introduce \emph{fail-stop} bugs.
For example, if a worker thread directly updates the text displayed on the UI thread, the app will crash.
Another example is, when an async thread finishes its background task, and tries to send a UI update event to a GUI component. Before the update takes effect, if the user rotates the screen, the UI thread will destroy and recreate that GUI component. By default, the update event is routed to the destroyed GUI rather than the newly created one, which may crash the app. 
In this paper, we term such fatal programming errors that violate the rules implied by the single-UI-thread model as \emph{async programming errors} (APEs).

Such bugs in Android are \emph{not easy to detect manually}, due to (1) they usually reside in the code of handling interactions between UI thread and async threads, which can be rather complicated for manual analysis; (2) they can only be triggered at the right states of GUI components (\eg, activity, fragment) with complicated lifecycle~\cite{activity,fragment}; (3) they have to be triggered at right thread scheduling, while the execution time of async threads is affected by the task and its running environment (\eg, network stability, system load). 

Even worse, \emph{existing bug detection techniques are ineffective for such bugs}. First, most GUI testing techniques, \eg, random testing~\cite{Monkey,MachiryTN13}, search-based testing~\cite{MahmoodMM14,MaoHJ16}, and model-based testing~\cite{AmalfitanoFTCM12,YangPX13,AzimN13,AmalfitanoFTTM15,Stoat2017}, are designed for functional testing in general. They aim at enumerating all possible event sequences (GUI-level events in particular) to manifest bugs, which is unscalable and time-consuming. 
Additionally, they mainly aim at improving code coverage, which may not be sufficient for exhibiting APEs --- require specific event sequences with appropriate lifecycle states and thread scheduling.
Second, static analysis tools, \eg, Lint~\cite{lint}, FindBugs~\cite{findbugs} and PMD~\cite{pmd}, although scalable, only enforce simple rules (syntax or trivial control/data-flow analysis) to locate suspicious bugs. For example, Lint declares it can find ``WrongThread" errors (one type of APEs)~\cite{lintchecks}.
However, as our evaluation in Section~\ref{sec:eval} demonstrates, Lint incurs a number of false negatives --- failing to detect those sophisticated ``WrongThread" errors as well as other types of APEs. Third, other fault detection techniques~\cite{Zhang2012,HsiaoPYPNCKF14,MaiyaKM14,Bielik15} (\eg, data race detection) have only tackled parts of APEs.
To systematically tackle APEs, we conducted a formative study on 2097 Android apps to understand them. First, we find the async constructs are indeed widely used in 48.6\% apps, and \textsf{\small{AysncTask}}s and \textsf{\small{Thread}}s account for the majority. Second, we identified 3 async programming rules (see Section~\ref{sec:study}) implied by the single-GUI-thread model by analyzing Android docs, technical posts and previous fault studies on async programming. Third, from 1019 apps that use async constructs, we found that developers do violate these rules and introduce APEs. We collected 375 real APEs, involving 9 exception types (thrown from apps), \eg, \textsf{\small{CalledFromWrongThread}}, \textsf{\small{IllegalStateException}}, \textsf{\small{BadTokenException}}, \etc.

Informed by the above results, we develop an approach \tool.
It first characterizes 3 fault patterns from 375 issues based on the 3 rules, and synergistically combines static analysis and dynamic UI exploration to efficiently manifest APEs. Specifically, it encodes the fault patterns into a static analyzer, locates suspicious APEs in the app code, generates a set of program paths that can reach the faulty code, maps program traces to real event sequences with appropriate environment, and finally verify these errors on the app.

We evaluate \tool on a set of 40 real-world Android apps, and compare it with three state-of-the-art GUI testing tools (Monkey~\cite{Monkey}, Sapienz~\cite{MaoHJ16}, and Stoat~\cite{Stoat2017}), and two fault detection tools (Lint~\cite{lint} and EventRacer~\cite{Bielik15}). 
The results show (1) \tool unveils and successfully processes 61 APEs, of which 51 can be reproduced (83.6\% hit rate) with real tests;
(2) \tool detects 3X more APEs than the testing tools, and reduces detection time from half an hour to a few minutes; and (3) within comparable analysis time, \tool detects 5X more APEs than EventRacer with very few false positives.

To summarize, this paper makes the following contributions:
\begin{itemize}[leftmargin=*]
\item We conduct a formative study on 2097 Android apps to investigate APEs, and identify three async programming rules implied by the single-GUI-thread model.

\item We develop \tool, a technique that synergistically combines static analysis and dynamic UI exploration, to efficiently detect and verify APEs.

\item We evaluate \tool on 40 real-world apps, and clearly demonstrate its effectiveness over the state-of-the-art testing and other fault detection techniques on APEs.
\end{itemize}

%% file: overview.tex
\vspace{-2mm}
\section{Async Programming Errors}
\label{sec:overview}

\subsection{Async Programming in Android}
\label{sec:async}

\begin{figure}[t]
	\centering
	\includegraphics[width=0.4\textwidth]{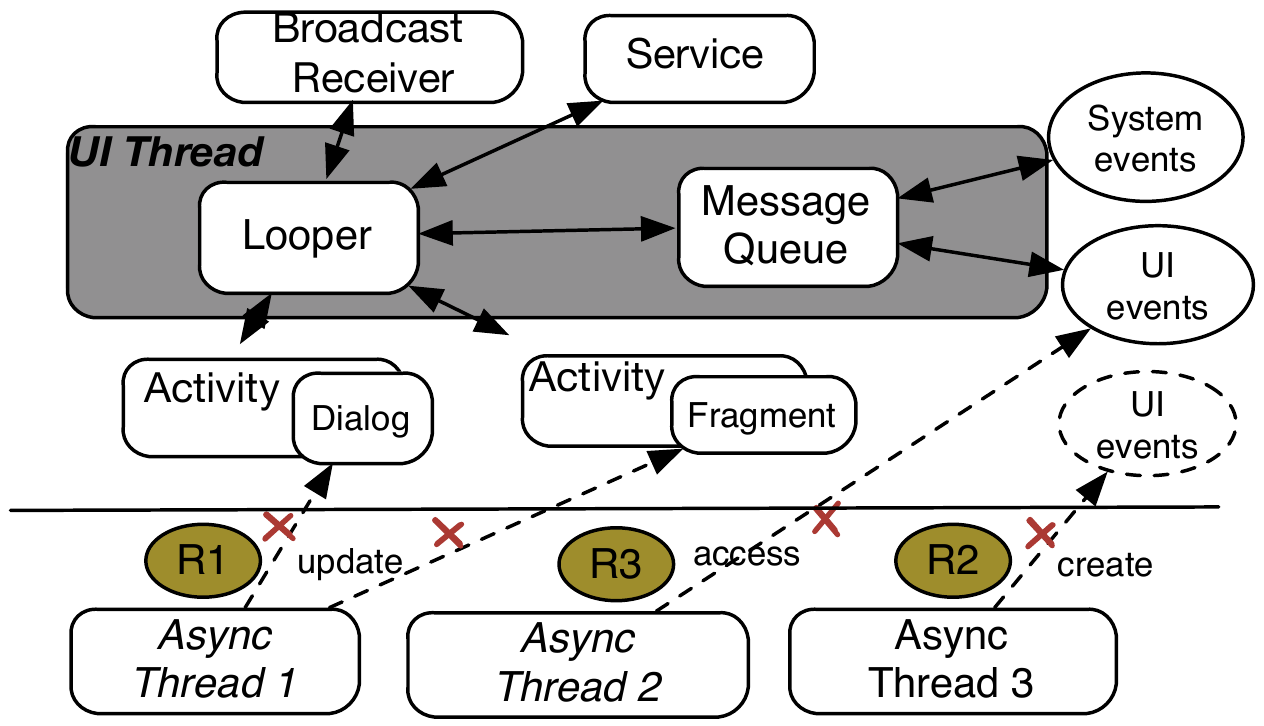}
	\vspace{-2mm}
	\caption{Single-GUI-thread model of Android and its three basic rules.}
	\label{fig:single_gui_thread}
	\vspace{-6mm}
\end{figure}

Fig.~\ref{fig:single_gui_thread} depicts the single-GUI-thread model of Android. The UI thread maintains a \textsf{\small{MessageQueue}}, and its \textsf{\small{Handler}} enqueues system or UI events into this queue. These events come from app components (\eg, \textsf{\small{Broadcast Receiver}}s, \textsf{\small{Service}}s) or GUIs (\eg, \textsf{\small{Activity}} and its associated visible components like \textsf{\small{Dialog}}s or \textsf{\small{Fragment}}s).
The UI thread's \textsf{\small{Looper}} dequeues events in a sequential order and dispatches them to the \textsf{\small{Handler}} for processing. The UI thread invokes the corresponding event handler \wrt an event and updates GUIs if necessary.

\begin{figure}[t]
	\centering
	\includegraphics[width=0.52\textwidth]{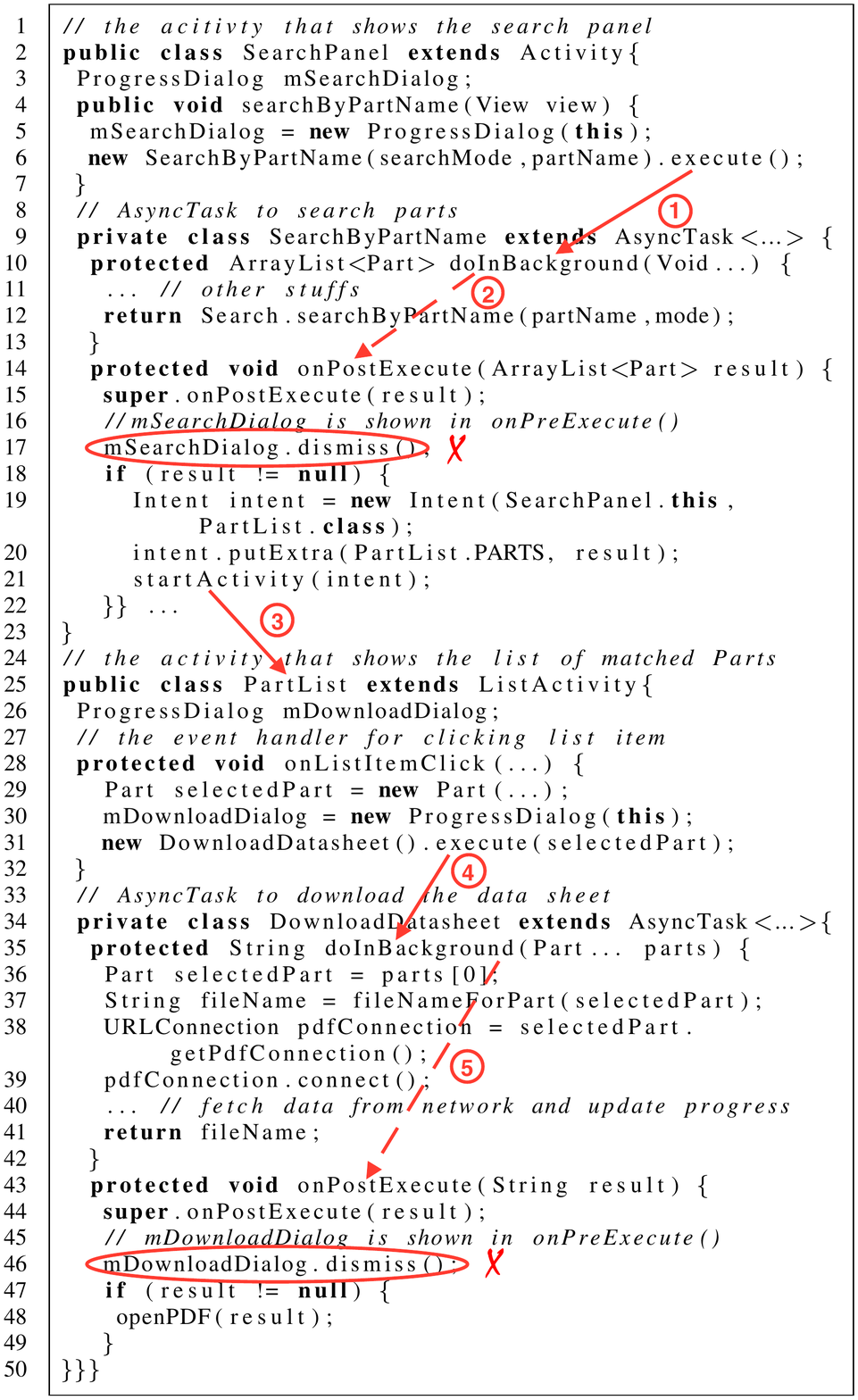}
	\vspace{-6mm}
	\caption{Motivating Example}
	\label{fig:adsdroid}
	\vspace{-5mm}
\end{figure}

ADF provides various async programming constructs~\cite{android_doc}. There are four typical constructs, \ie, \textsf{\small{AysncTask}}, \textsf{\small{Thread}}, \textsf{\small{AsyncTaskLoader}} and \textsf{\small{IntentService}}. Among them, \textsf{\small{AsyncTask}} allows one to perform short background operations and publish results on the UI thread; \textsf{\small{Thread}} is inherited from Java, and executes tasks in the background; \textsf{\small{AsyncTaskLoader}} utilizes \textsf{\small{AysncTask}} to perform async data loading, and has similar callbacks as \textsf{\small{AsyncTask}}, but it is lifecyle aware: ADF binds/unbinds the worker thread according to GUI's lifecyle.  \textsf{\small{IntentService}} handles async requests in an async thread, and sends the results to the UI thread vis a \textsf{\small{Broadcast Receiver}}.
Fig.~\ref{fig:adsdroid} illustrates the use of \textsf{\small{AysncTask}}s in \emph{ADSdroid}~\cite{adsdroid}, it starts two async threads, \ie, \textsf{\small{SearchByPartName}} (Lines 9-22) and \textsf{\small{DownloadDatasheet}} (Lines 34-49) to search electronic components' datasheet, and download from a remote server if requested. 
The activity \textsf{\small{SearchPanel}} (Lines 2-23) searches for the result with user input in the \textsf{\small{doInBackground}}, showing a progress dialog in \textsf{\small{onPreExecute}} before searching, and dismisses it via \textsf{\small{onPostExecute}}.
The results are shown in a \textsf{\small{ListView}} of the activity \textsf{\small{PartList}} (Lines 25-50), in which users can click any matched item (Lines 28-32) for downloading.

However, there are two APEs (Lines 17 and 46), neither of which has been covered by developers. The root causes for these two APEs are similar. When users rotate the screen right after the start of the aysnc tasks \textsf{\small{SearchByPartName}} and \textsf{\small{DownloadDatasheet}} but before they finish (\ie, before the execution of \textsf{\small{onPostExecute}}s), the app will crash when \textsf{\small{mSearchDialog}} (Line 17) and \textsf{\small{mDownloadDialog}} (Line 46) are dismissed.
Because the rotation will destroy the current
activity and create a new one, but the dialogs were attached to the
original one, which does not exist anymore.
This leads to a fatal \textsf{\small{BadTokenException}}.
We can see the right timing and lifecycle states are crucial to manifest these APEs.
This paper aims at tackling such hard-to-detect errors.

\vspace{-2mm}
\subsection{Formative Study}
\label{sec:study}
To understand APEs, we conducted a formative study to investigate the following questions. This enables our problem definition in Section~\ref{sec:problem_definition} and fault pattern analysis in Section~\ref{sec:fault_pattern_analysis}.

\noindent\emph{\textbf{Q1: Are async constructs widely-used by Android developers to follow the single-GUI-thread model?}}
To answer this, we focus on the four typical constructs introduced in Section~\ref{sec:async}, and investigate 2097 open-source Android apps from F-droid to observe the use of these constructs since F-droid is one of largest Android app repositories, which covers diverse application categories; and all apps are open-source and maintained on Github, Google Code, \etc. We use Soot~\cite{soot}, a static analysis tool, to identify aysnc constructs.

\noindent\emph{\textbf{Answer:}} Async constructs are widely-used by Android developers. We found 1019 out of 2097 apps use async constructs (account for 48.6\%). 
Among 1019 apps, 2968 \textsf{\small{AsyncTask}}s, 1248 \textsf{\small{Thread}}s, 286 \textsf{\small{IntentService}}s, and 35 \textsf{\small{AsyncTaskLoader}}s are used. \textsf{\small{AsyncTask}} and \textsf{\small{Thread}} account for the majority, which also conforms to prior work~\cite{Lin14,LinOD15}. 

\begin{figure*}[t]
	\centering
	\includegraphics[width=0.9\textwidth]{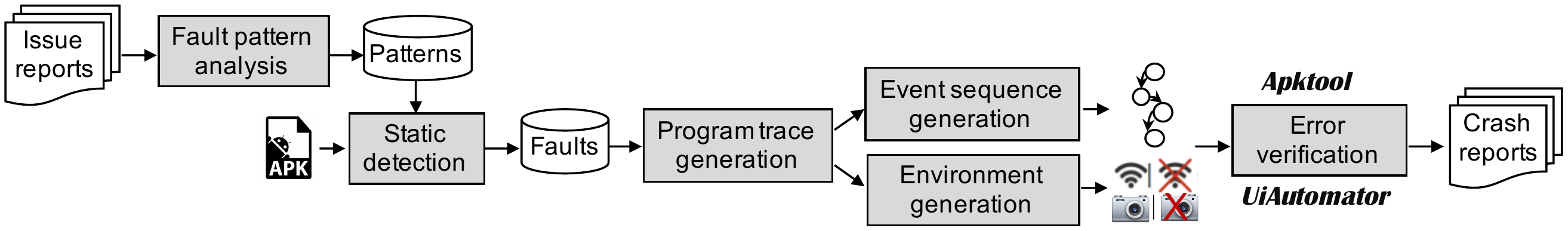}
	\vspace{-3mm}
	\caption{Workflow of \tool}
	\label{fig:overview}
	\vspace{-2mm}
\end{figure*}

\noindent\emph{\textbf{Q2: Are there any basic rules implied by the single-GUI-thread model that Android developers should obey in async programming? Are they common to other GUI frameworks?}}
To answer this, we conducted a thorough and careful inspection on (1) Android docs and APIs~\cite{android_doc}, including the principle of single-GUI-thread model~\cite{thread}, various async programming constructs~\cite{asynctask,thread1,handler,looper,loader,intentservice,handlerthread,threadpool}, GUI components~\cite{activity,fragment,toast,dialog,listview}, \etc; and (2) technical posts filtered from Stack Overflow (the largest developer Q\&A community) by the keywords ``Android'' plus the names of async constructs, tutorials on async programming~\cite{codepath}; and (3) fault studies on Android~\cite{Hu11,Zaeem14,Linares2017,fan18}.

\noindent\emph{\textbf{Answer:}} We identified 3 async programming rules (Fig.~\ref{fig:single_gui_thread} annotated these rules as \emph{R1}, \emph{R2} and \emph{R3}), which are also common to other modern GUI frameworks.

\noindent \emph{\textbf{$\bigcdot$ Rule 1 (Async threads should not update GUI objects)}}: Since Android UI toolkit is not thread-safe, the single-GUI-thread model requires that async threads should not directly manipulate GUI objects. GUI objects include \emph{visible components} (\eg, update a \textsf{\small{Dialog}}'s message) and \emph{data models} (\eg, change the content of \textsf{\small{ArrayAdapter}} that fits in between a \textsf{\small{ListView}} and an \textsf{\small{ArrayList}}). Instead, they should designate the UI thread to handle GUI objects via UI-safe methods like \textsf{\small{Activity\#runOnUiThread}}~\cite{thread}. This rule is also enforced by many GUI frameworks, \eg, \textsf{\small{Swing}}, \textsf{\small{SWT}}, \textsf{\small{Qt}}, and \textsf{\small{Cocoa}}.

\noindent \emph{\textbf{$\bigcdot$ Rule 2 (Async threads should not create GUI components in the background)}}:
The UI thread by default is created with a \textsf{\small{Handler}} and a \textsf{\small{Looper}}. The \textsf{\small{Handler}} enqueues events (\eg, messages or runnable objects) into a message queue. These events come from different app components and GUIs. The \textsf{\small{Looper}} dequeues and dispatches the events to the \textsf{\small{Handler}} for processing.
However, an async thread (except \textsf{\small{HandlerThread}}) by default is not associated with a \textsf{\small{Handler}}, thereby it should not directly create GUI components (\eg, \textsf{\small{Toast\#show}}, \textsf{\small{Dialog\#create}}) in the background.
Instead, they should post GUI creations via UI thread's \textsf{\small{Handler}} into \textsf{\small{Looper}} for processing. \textsf{\small{Qt}}'s  \textsf{\small{event loop}} and \textsf{\small{wxErlang}}'s \textsf{\small{mailbox queue}} enforce this similar rule. 

\noindent \emph{\textbf{$\bigcdot$ Rule 3 (Async threads should avoid accessing GUIs or performing transactions inside async callbacks)}}: Async callbacks such as \textsf{\small onPostExecute} run on the UI thread, but they have no knowledge of the current states of GUIs. Because they are called when the async thread returns. As a result, accessing GUIs or performing transactions for \textsf{\small{Fragment}}s in async callbacks (\eg, \textsf{\small{AysncTask\#onPost-\\Execute}}, \textsf{\small{LoaderCallbacks\#onLoadFinished}}) has the risks of sending GUI updates to destroyed GUIs and losing app state~\cite{stateloss}. Cocoa Touch (GUI framework of iOS) also enforces this similar rule~\cite{ios}.

\noindent\emph{\textbf{Q3: Do developers violate these rules? Are there any challenges to solve such APEs?}}
To answer this, we utilize Github and Google Code APIs to scrawl the issue reports of 1091 apps that uses async constructs in \textbf{\emph{Q1}}. To identify APEs, we only collect issues that are reported with exception traces, which contain the callbacks of async constructs; then inspect their issue descriptions, comments, patches if available to confirm valid issues that do violate the three rules.

\noindent\emph{\textbf{Answer:}} We finally got 375 valid APE issues. The number of valid issues is not large since many issues are reported without exception traces. It is reported only 16\% Github issues have exception traces~\cite{fan18}. But our evaluation (Section~\ref{sec:eval}) reveals APEs affect more than 25\% apps. These issues violate the rules in different forms, involving 9 different types of fatal exceptions, \eg, \textsf{\small{CalledFromWrongThread}}, \textsf{\small{IllegalStateException}}, and \textsf{\small{BadTokenException}}. 
The violations of \emph{Rule 1} and \emph{3} account for the majority in these 375 issues (91.2 \%).
By investigating the fixing process of these APEs on Github, we note developers face three main challenges: (1) Due to lack of adequate understanding, they usually use simple try-catches to fix APEs; (2) Many APEs reside in the releases since developers fail to discover them during development; (3) Due to the lack of reproducing tests, several developers complain about the difficulties of debugging.

\vspace{-2mm}
\subsection{Problem Definition}
\label{sec:problem_definition}
We name the errors that violate the three basic rules as \emph{aysnc programming errors}, and formulate our problem as follows.

\noindent{\textbf{UI thread.}} A UI thread $u$ is the main thread that is created when an app starts, and manages GUI components.

\noindent{\textbf{Async thread.}} An async thread $w$ is a worker thread that is instantiated from an async construct (\eg, \textsf{\small{Thread}}), started by the UI thread (\eg, \textsf{\small{Thread\#run}}), and executes a task. 

\noindent{\textbf{UI-accessing and UI-safe methods.}} A UI-accessing method (i.e., $m_{uiaccess}$) may create a GUI component (\eg, \textsf{\small{Toast\#show}}), change a GUI component's state (\eg, \textsf{\small{Dialog\#dismiss}}), or commit fragment transactions. A UI-safe method $m_{uisafe}$ permits safe GUI access, \eg, \textsf{\small{Activity\#runOnUiThread}}, \textsf{\small{Handler\#post}}, or GUI state check, \eg, \textsf{\small{Activity\#isFinishing()}}.

\noindent{\textbf{Program trace and its event sequence.}} A program trace $t$ is a sequence of method calls, $m_1, \ldots, m_i, \ldots, m_n$, on the call graph of an app. The event sequence $s$ \wrt $t$ is a set of user events, $e_1, \ldots, e_i, \ldots, e_k$ ($k \leq s$ usually holds), that can execute out $t$ under the given environment $E$ (\eg, thread scheduling, network status, app permissions, sensor inputs and platform versions).

\noindent{\textbf{Problem Definition.}}
Our problem is to check whether there exists any trace $t$ on which $u$ creates an async thread $w$, and $w$ invokes any UI-accessing method that is \emph{not control-dependent} on a UI-safe method. If $t$ exists, we find an event sequence $l$ that follows $t$ to exhibit the error under given environment $E$.

\vspace{-2mm}
\subsection{Prior Work}
\label{sec: prior_work}
Prior work has only tackled parts of this problem.
One close work is from Zhang \etal~\cite{Zhang2012}, which finds invalid thread access errors in Java GUI applications (\textsf{\small{Swing}}, \textsf{\small{SWT}}, \textsf{\small{Android}}), and gives warnings in the form of method call chains. However, this work has several significant differences from ours: First, they only handle a subset of errors \wrt \emph{Rule 1}, \ie, an async thread directly accesses GUI components.
Second, they use static analysis to check whether a thread spawning can reach a GUI object accessing method, which is determined by the \textsf{\small{ViewRoot\#checkThread}} method in ADF. This is ad hoc and may miss many errors. Third, they do not confirm errors, which may bring many false positives. In contrast, we conduct a systematic study to understand APEs, and combine static and dynamic analysis to confirm errors with real tests.
Lin \etal~\cite{Lin14} investigate the uses of \textsf{\small{AsyncTask}}, and observe the invalid thread access errors. But they aim to solve another different problem of automatically refactoring long-running tasks in UI thread into \textsf{\small{AysncTask}}s.

Another area of related work is detecting and reproducing concurrency bugs (data race in particular)~\cite{Lin14,HsiaoPYPNCKF14,MaiyaKM14,Bielik15,OzkanET15,Tang2016,Li0GXMML16}. 
They treat some APEs \wrt \emph{Rule 3} as data races. For example, the two APEs in Fig.~\ref{fig:adsdroid} can be interpreted as data races since the \textsf{\small{mPanels}} variable inside the framework class \textsf{\small{PhoneWindow}} can be accessed simultaneously by the UI and async thread in which one access is a write operation.
Existing data race detection tools~\cite{HsiaoPYPNCKF14,MaiyaKM14,Bielik15} exploit dynamically explored event traces to build a happen-before graph, and then query the graph to find potential data races.
But they have several limitations in practice to detect APEs: First, their effectiveness heavily relies on the traces for building the graph. If the traces have not fully covered app code, the detection ability is limited. Second, they usually generate a large number of false positives due to conservative analysis~\cite{Hu2016}.
Existing data race reproducing tools~\cite{OzkanET15,Tang2016} are also impractical since they either require user-provided traces or depend on the results of data race detection tools when confirming bugs. 
Section~\ref{sec:eval} compares \tool with existing data race detection tools to confirm these observations.

%% file: approach.tex
\section{Our Approach \tool}
\label{sec:approach}

This section details our approach \tool. Figure~\ref{fig:overview} shows its workflow, which is composed of five key steps: \tool (1) summarizes a set of fault patterns from the collected APE issues, (2) encodes these fault patterns into a static analyzer to locate faulty code, (3) generates a set of program traces that can reach the faulty code from the app entry, (4) maps the program traces into real event sequences (tests) with appropriate environment; and (5) verifies APEs with the tests, and dumps crash reports for fixing.

\subsection{Fault Pattern Analysis}
\label{sec:fault_pattern_analysis}

We categorize 375 APE issues into three groups \wrt the rules summarized in the formative study. By analyzing these issues, we characterize these APEs as three fault patterns.

\begin{figure}[t]
	\centering
	\includegraphics[width=0.4\textwidth]{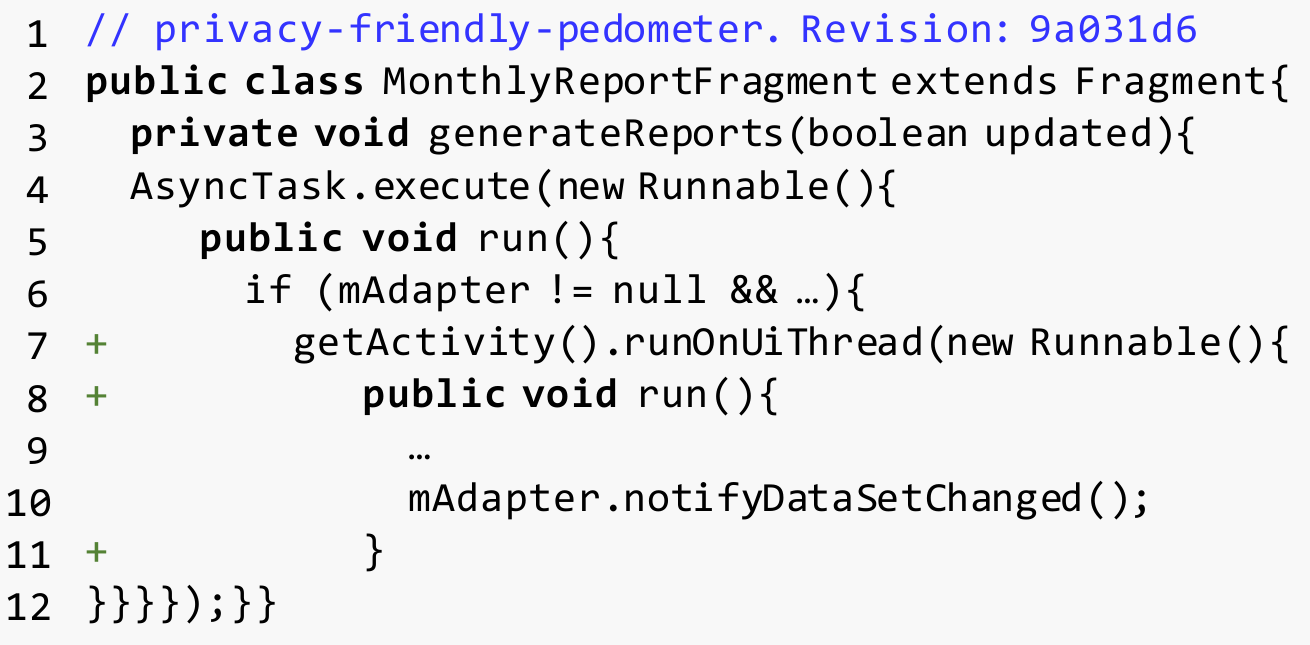}
	\vspace{-3mm}
	\caption{Example of Fault Pattern 1}
	\label{fig:uiupdate}
	\vspace{-1mm}
\end{figure}

\noindent \emph{\textbf{$\bigcdot$ Fault Pattern 1:}} If an async thread $w$ is started, and $w$ calls a UI-access method $m_{uiaccess}$, which is not control-dependent on a UI-safe method $m_{uisafe}$.
This pattern violates \emph{\textbf{Rule 1}}, which is represented as
\emph{start}($w$) $\rightarrow$ $\neg$ $m_{uisafe}$ $\rightarrow$ $m_{uiaccess}$.
Fig.~\ref{fig:uiupdate} shows such an error of \emph{Pedometer}~\cite{pedometer}, an app that stores the user's step count per hour (the symbol ``+'' denotes the corresponding patch to fix this error).
In fragment \textsf{\small{MonthlyReportFragment}}, it starts an aysnc thread to generate the monthly report and refresh the GUI by invoking \textsf{\small{notifyDataSetChanged}}, which crashes the app.

\begin{figure}[t]
	\centering
	\includegraphics[width=0.4\textwidth]{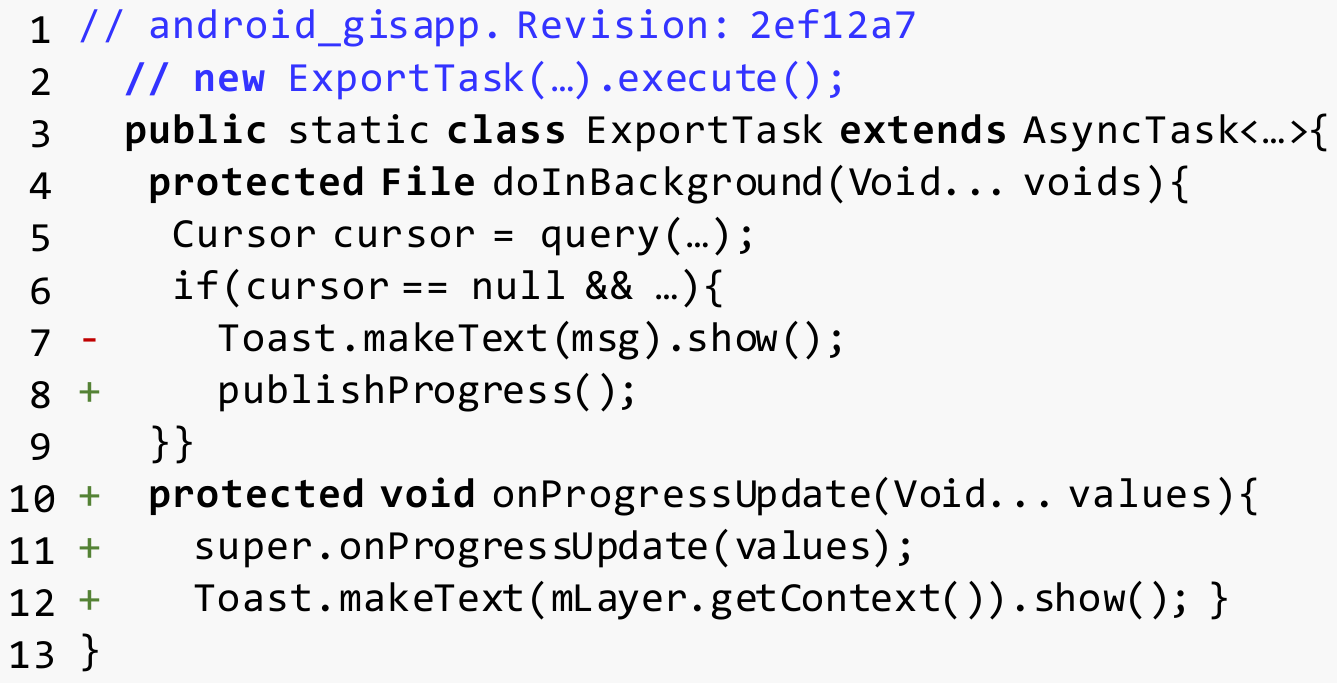}
	\vspace{-3mm}
	\caption{Example of Fault Pattern 2}
	\label{fig:looper}
	\vspace{-3mm}
\end{figure}

\noindent \emph{\textbf{$\bigcdot$ Fault Pattern 2:}} If an async thread $w$ is started, and $w$ calls a GUI creation method $m_{uicreate}$, which is not posted on the UI thread by $m_{postlooper}$ to execute.
This pattern violates \emph{\textbf{Rule 2}}, which is represented as
\emph{start}($w$) $\rightarrow$ $\neg$ $m_{postlooper}$ $\rightarrow$  $m_{uicreate}$. 
Fig.~\ref{fig:looper} shows such an error of \emph{gisapp}~\cite{gisapp}, which is a user interface controls library for Android geo applications (the symbols ``+'' and ``-'' denote the corresponding patch to fix this error).
\emph{gisapp} uses an async thread \textsf{\small{ExportTask}} to export the data by retrieving the database, and use \textsf{\small{Toast\#makeText}} to show a message if no data is available. However, \textsf{\small{ExportTask}} has not referred to the UI thread's \textsf{\small{Handler}} to post messages, which crashes the app.

\noindent \emph{\textbf{$\bigcdot$ Fault Pattern 3:}} 
If an async thread $w$ is started, and before $w$ returns, the target activity or fragment $A$ is destroyed or stopped, and after $w$ returns, its async callback calls a UI-access method $m_{uiacess}$, which is not control-dependent on a UI-safe method $m_{uisafe}$.
This pattern violates \emph{\textbf{Rule 3}}, which is represented as
\emph{start}($w$) $\rightarrow$ \emph{destroy}($A$) or \emph{stop}($A$) $\rightarrow$ \emph{return}($w$) $\rightarrow$ $\neg$ $m_{uisafe}$ $\rightarrow$ $m_{uiaccess}$.
In Fig.~\ref{fig:adsdroid}, there are two such errors. When users rotate the screen right after the start of the \textsf{\small{AsyncTask}}s \textsf{\small{SearchByPartName}} and \textsf{\small{DownloadDatasheet}} but before they finish,  the dismiss of \textsf{\small{mSearchDialog}} (Line 17) and \textsf{\small{mDownloadDialog}} (Line 46) can crash the app. 

\subsection{Static Fault Detection}
\label{sec:fault_pattern}

Algorithm~\ref{algo:fault_detection} details static fault detection. It takes as input an app, the lists of UI-accessing and UI-safe methods (summarized from Android docs), and outputs the APE locations (including the starting points of async threads). 
Algorithm~\ref{algo:fault_detection} works on the static call graph of an app, where each node denotes a method, and each edge $\langle f,g \rangle$ denotes the call from the method $f$ to $g$.
We first identify all method nodes that start async threads (Line 3) and check whether the use of corresponding async constructs violate the rules (Lines 4-17). 
For each aysnc construct, we get the overridden methods according to its class type (Lines 5-6). Take \textsf{\small{AsyncTask}} as an example, we consider its callback methods \textsf{\small{onPreExecute}}, \textsf{\small{doInBackground}} and \textsf{\small{onPostExecute}}, \etc. By querying the call graph, we gets all the nodes called by these methods (Line 7).
If a UI-accessing method node is not control-dependent on any UI-safe methods (\eg, \textsf{\small{Activity\#runOnUiThread}}), this node will be tagged as suspicious APE (Lines 12-16). Note that we conduct both intra- and inter-procedural control-dependent analysis to reduce false alarms.
Algorithm~\ref{algo:fault_detection} also checks the successor methods that are called by this node (Line 17) to avoid false negatives.

\SetKwInput{KwInput}{Input}
\SetKwInput{KwInput}{Input}
\SetKw{Let}{let}
\SetKw{Continue}{continue}

\begin{algorithm2e}[t]
	\footnotesize
	\setcounter{AlgoLine}{0}
	\caption{Static Fault Detection}
	\label{algo:fault_detection}
	\DontPrintSemicolon
	\SetCommentSty{mycommfont}
	\KwIn{$apk$: an Android app, $uiAccessAPIs$: list of UI-access methods, $uiSafeAPIs$: list of UI-safe methods}
	\KwOut{$APEs$: List$<$methodStartThread, stmtStartThread, methodAccessUI, stmtAccessUI$>$}
	
	$APEs$ $\gets$ $\emptyset$  \;
	$cg$ $\gets$ buildCallGraph($apk$) \;
	$asyncStartNodes$ $\gets$ getReachableAsyncStarts($cg$) \;
	\ForEach{Node $node$ $\in$ $asyncStartNodes$}{
		$async$ $\gets$ getAasyncClass($node$)\;
		$o$ $\gets$ getOverriddenMethods($async$) \;
		$nodeList$ $\gets$ getCalledMethodNodes($o$, $cg$) \;
		\While{$nodeList \neq \emptyset$}{
			$node \gets$ $nodeList$.dequeue() \;
			\If{isVisited($node$)}{ 
				\Continue
			}
			\If{$node$ $\in$ $uiAccessAPIs$}{ 
				\If{$\neg$ controlDepdent($node$, $uiSafeAPIs$) }{
					$m$ $\gets$ getCallingMethod($async$)\;
					$n$ $\gets$ getCallingMethod($node$)\;
					$APEs$ $\gets$ $\left \langle m, async, n, node \right \rangle$
				}
			}
			$nodeList$.enqueue(getCalledMethodNodes($node$, $cg$))
		}
	}
	\Return $APEs$
\end{algorithm2e}

\noindent\emph{\textbf{Example}}. In Fig.~\ref{fig:adsdroid}, \tool identifies that \textsf{\small{searchByPartName}} starts an \textsf{\small{AsyncTask}}, \textsf{\small{SearchByPartName}}. In the \textsf{\small{onPostExecute}} callback, it locates a UI-accessing method \textsf{\small{ProgressDialog\#dismiss}} (Line 17), which is not control-dependent on any UI-safe methods, \eg, \textsf{\small{Activity\#isFinishing()}}, to safely check activity state. So \tool reports a suspicious APE.

\subsection{Program Trace Generation}
\label{sec:backward_trace}

Algorithm~\ref{algo:backward} details the program trace generation (the idea is similar to backward symbolic execution~\cite{MaKFH11,SuFPHS15}). It takes as input a target method and statement (\ie, \textsf{\small{methodAccessUI}} and \textsf{\small{stmtAccessUI}} from Algorithm~\ref{algo:fault_detection}), and returns the traces (\ie, method call sequences) that can reach the APE.
It starts with an empty trace $t$, and backtracks to reach the entry activity (Lines 5-30). Specifically, the maximum number of generated traces (\textsf{\small{MaxTraceCnt}}, Line 17) and the maximum trace length ( \textsf{\small{MaxTraceLen}}, Line 5) is configurable.

During the analysis, each trace $t$ may have one of the three states, \ie, \textsf{\small{pending}} ($t$ is under propagation), \textsf{\small{terminated}} ($t$ reaches the entry activity), \textsf{\small{failed}} (the propagation fails due to dead code or limitations of static analysis tools). If no traces are pending, the analysis stops (Lines 6-7). Otherwise, the analysis continues until the maximum trace length is reached.
At each iteration, the algorithm uses two important variables, \ie, \textsf{\small{ptrMethod}} and \textsf{\small{ptrStmt}}, pointing to the current backtrack point of $t$.
When $t$ is pending, the algorithm queries the callers of $t$ via \textsf{\small{getAcyclicCallers}} (we details this function later). If there are no callers, $t$ will be set as \emph{failed} due to the propagation cannot proceed (Lines 12-14).
Otherwise, we update \textsf{\small{ptrMethod}} and \textsf{\small{ptrStmt}} of $t$ (Lines 16-26). Specifically, if $t$'s \textsf{\small{ptrMethod}} has multiple callers, the algorithm will fork out a new trace $t'$ from $t$ (copying all previous backtrack information from $t$ to $t'$), and add $t'$ into the list of traces (Lines 16-22). 

Here, \textsf{\small{updateTrace}} (Line 21 and 25) performs three operations: (1) check whether $t$'s \textsf{\small{ptrMethod}} has reached the entry activity and update $t$'s state accordingly. In particular, if \textsf{\small{ptrMethod}} is a callback method (\eg, user event handler) of the entry activity, $t$ will be set as \textsf{\small{terminated}}. (2) update the method call chain of $t$ by adding \textsf{\small{ptrMethod}}. The method call chain will be later used to generate event sequences. (3) record the conditions in the body of \textsf{\small{ptrMethod}} that \textsf{\small{ptrStmt}} are control-dependent on. These conditions will be analyzed later to create necessary environment to improve the hit rate. Section~\ref{sec:event_sequence} will discuss the details of (2) and (3).

The function \textsf{\small{getAcyclicCallers}} queries and returns the immediate callers of $t$'s \textsf{\small{ptrMethod}}. It removes the visited callers to ensure the traces are acyclic. When backtrack, we differentiate three types of call relations that widely exist in Android.

\noindent \emph{\textbf{$\bigcdot$ Explicit Calls}}. The function $f$ immediately calls function $g$, \ie, there exists an edge $\langle f,g \rangle$ on the call graph. For example, \textsf{\small{SearchByPartName\#doInbackground}} is explicitly called by \emph{searchByPartName} via \textsf{\small{AsyncTask\#execute}} (see Fig.~\ref{fig:adsdroid}).

\noindent \emph{\textbf{$\bigcdot$ Implicit Calls}}. Android systems have many implicit calls through its framework. An implicit call from function $f$ to $g$ indicates $f$ calls $g$ asynchronously. For example, the callback \textsf{\small{onPostExecute}} of \textsf{\small{SearchByPartName}} is implicitly called when the callback \textsf{\small{doInbackground}} returns (see Fig.~\ref{fig:adsdroid}).

\noindent \emph{\textbf{$\bigcdot$ Inter-Component Transition Calls}}. Android system uses \textsf{\small{Intent}}s to start activities or services. For example, in \textsf{\small{SearchByPartName}}, the activity \emph{PartList} is called by the callback \textsf{\small{onPostExecute}} via \textsf{\small{startActivity}} (see Fig.~\ref{fig:adsdroid}).

\noindent\emph{\textbf{Example.}} We explain Algorithm~\ref{algo:backward} on \emph{ADSdroid} (Fig.~\ref{fig:adsdroid}). Algorithm~\ref{algo:fault_detection} reports two suspicious faults, one of which is in the \textsf{\small{onPostExecute}} callback, which dismisses \textsf{\small{mDownloadDialog}} without any UI-safe methods (Line 46). Algorithm~\ref{algo:backward} first initializes a pending trace, whose \textsf{\small{ptrMethod}} is set as \textsf{\small{onPostExecute}} and \textsf{\small{ptrStmt}} as the dismiss statement. It then finds \textsf{\small{doInBackground}} is an implicit caller of \textsf{\small{onPostExecute}} (see \textcircled{5}). Next, it finds \textsf{\small{onListItemClick}} explicitly calls \textsf{\small{doInBackground}} via \textsf{\small{execute}} 
(see \textcircled{4}).
Next, it finds the activity \textsf{\small{PartList}} is called by \textsf{\small{SearchByPartName\#onPostExecute}} (see \textcircled{3}), and update \textsf{\small{ptrMethod}} as \textsf{\small{onPostExecute}} and \textsf{\small{ptrStmt}} as the \textsf{\small{startActivity}} statement (Line 21). Similarly, it finally finds the method \textsf{\small{searchByPartName}} starts the \textsf{\small{SearchByPartName}} (see \textcircled{1}). The final trace is \textsf{\small{searchByPartName}} $\rightarrow$ \textsf{\small{onListItemClick}} (only user event handlers are shown).

\SetKwInput{KwInput}{Input}
\SetKwInput{KwInput}{Input}
\SetKw{Let}{let}
\SetKw{Continue}{continue}
\begin{algorithm2e}[t]
\footnotesize
\setcounter{AlgoLine}{0}
\caption{Program Trace Generation}\label{algo:backward}
\DontPrintSemicolon

\KwIn{$method$: the target method, $stmt$: the target statement}
\KwOut{$traces$: the list of candidate traces}

\BlankLine

\Let $t$ be an initially empty trace\; 
\Let $traceLen$ be the current trace length (initialized as 0)\;
$t$.ptrMethod $\gets$ $method$, $t$.ptrStmt $\gets$ $stmt$ \;
$traces$ $\gets$ \{$t$\} \;

\While{$traceLen$ $<$ MaxTraceLen}{
	\If{ $\neg$ hasPendingTrace($traces$)}{
		\Return $traces$
	}
	$new\_traces$ $\gets$ \{\}\;
	\ForEach{Trace $t \in traces$}{
		\If{ isPending($t$)}{\tcp*[h]{$t$ is a pending trace}\;
			$t$.callers $\gets$ getAcyclicCallers($t$.ptrMethod)\;
			\If{t.callers == $\emptyset$}{
				\tcp*[h]{set $t$'s state as failed}\;
				updateTrace($t$)\;
				\Continue 
			}
			\Let $c$ be the first caller of $t$.callers\;
			\ForEach{Caller $c'$ $\in$ $t$.callers$ \setminus$\{c\}}{
				\If{count($traces$) $<$ MaxTraceCnt}{
					$t'$ $\gets$ fork($t$)\;
					$t'$.ptrMethod $\gets$ getMethod($c'$)\;
					$t'$.ptrStmt $\gets$ getCallsite($c'$)\;
					updateTrace($t'$)\;
					$new\_traces$ $\gets$ $new\_traces$ $\cup$ \{$t'$\}\;
				}
			}
			$t$.ptrMethod $\gets$ getMethod($c$)\;
			$t$.ptrStmt $\gets$ getCallsite($c$)\;
			updateTrace($t$)\;
			$new\_traces$ $\gets$ $new\_traces$ $\cup$ \{$t$\}\;
		}\Else{
			\tcp*[h]{$t$ is a terminated or failed trace}\;
			$new\_traces$ $\gets$ $new\_traces$ $\cup$ \{$t$\}\;
		}
	}
	$traces$ $\gets$ $new\_traces$\;
	$traceLen$ $\gets$ $traceLen$ + 1 \;
}
\end{algorithm2e}

\subsection{Event Sequence and Environment Generation}
\label{sec:event_sequence}

This step converts the program traces from Section~\ref{sec:backward_trace} into actionable event sequences with appropriate environment.

\noindent\emph{\textbf{Event Sequence Generation}}. \tool considers two main types of callbacks when converting a program trace.

\noindent\emph{\textbf{$\bigcdot$ User event handler callbacks.}} \tool maps each event handler on the trace to a corresponding action \wrt a UI widget or view. For example, in Fig.~\ref{fig:adsdroid}, the event handler \textsf{\small onListItemClick(...)} is mapped to a \textsf{\small click} action on a \textsf{\small{ListView}} item; \textsf{\small searchByPartName} (declared in the XML layout file) is mapped to a \textsf{\small click} action on the ``Search" button.
To achieve this, \tool maintains a view-handler mapping table, which supports event handler callbacks registered in both app code and XML layouts~\cite{RountevY14,Song2017}.

\noindent\emph{\textbf{$\bigcdot$ Activity/Fragment lifecycle callbacks.}}
For an activity or fragment lifecycle callbacks (\eg, \textsf{\small onRestart}, \textsf{\small onResume}, \textsf{\small onDestroy}) on the trace, \tool automatically substitutes the callback with special events that can force app to execute it. For example, for \textsf{\small onRestart}, \tool generates a long-press ``Home''  action (show the list of recently-opened apps) followed by a touch action on this app (switch back to the previous app again), triggering the lifecycle transition \textsf{\small onStop} $\rightarrow$ \textsf{\small onRestart}; for \textsf{\small onDestroy}, \tool generates a device-rotation event, triggering the transition \textsf{\small onStop} $\rightarrow$ \textsf{\small onDestroy} $\rightarrow$ \textsf{\small onStart}. 
Currently, \tool considers \textsf{\small Activity}, \textsf{\small Fragment} and other lifecycle-aware components (\eg, \textsf{\small Loader}).

\noindent\emph{\textbf{Environment Generation}}.
\tool automatically constructs appropriate environment for event sequences. During the trace generation, \tool records the control-dependent conditions of an APE, and analyzes them to derive the environment.
\tool currently focuses on three types of environment, which we find they can cover most of cases.

\noindent\emph{$\bigcdot$ Specific user inputs.} \tool tracks the constraints on user inputs that can affect error verification. It currently supports two lightweight strategies: (1) infer the required input formats (\eg, email address, phone number) from the property \textsf{\small android:inputType} in the UI layout files; and (2) track the intra-procedural data-flows of inputs and infer the required contents by analyzing simple string APIs (\eg, equal to a constant string, contain a specific character).

\noindent\emph{$\bigcdot$ Explicit system settings or permissions}. 
\tool analyzes specific APIs in the recorded conditions to infer necessary system settings or permissions. For example,
if $\neg$\textsf{\small WifiManager\#isWifiEnabled()} is the condition, \tool will disable WiFi before replaying the test; if \textsf{\small Camera\#open()} is the condition, \tool will grant the camera access permission at runtime when required. 

\noindent\emph{$\bigcdot$ Specific Exception Handling}. Some APEs reside in exception handling code, and they cannot be manifested without triggering the corresponding exception. For example, if the APE can be only reached by an \textsf{\small IOException} when the app fails to access a remote server via network, \tool will disconnect the network before replaying the test to simulate the exception. \tool only handles specific cases of \textsf{\small IOException} (\eg, cannot access network or files), but it can be extended to support other exceptions if required.

Currently, \tool supports limited environment (\eg, specific user inputs, network connection, camera access, file access) in our evaluated subjects, and does not consider external events (\eg, sensor inputs and intents).
The environment can be extended in the future by using more sophisticated techniques like symbolic analysis~\cite{Rasthofer2017} and exception handling~\cite{ZhangE14}.

\vspace{-6mm}
\subsection{Error Verification}
\label{sec:errorv}
To confirm an APE, we replay the generated event sequence with appropriate environment on the target app, and monitor the intended 9 exception types via Android Debugging Bridge (\textsf{\small adb logcat}).
For those APEs whose manifestations require specific activity/fragment state and thread scheduling (\emph{Fault Pattern 3} in particular), we  
instrument the original app $A$ to $A'$ by adding semaphore operations $P$ (waiting) and $V$ (release) at appropriate program locations. For a suspicious faulty statement $i$ in the callback of async thread $w$, we insert a $P$ operation (in the background callback) before $i$ to wait for a signal; at the right activity or fragment lifecycle callback on the UI thread $U$, we insert a $V$ operation to send the signal. As a result, we are able to control the thread scheduling and make sure the scheduling happen at the right lifecycle state. 
Note that if an app correctly follows the 3 async programming rules, the control of thread scheduling will not introduce APEs or force the app to crash. The instrumentation method only amplifies the possibility of the long execution time to simulate real possible scenarios (\eg, delay of network access, wait of data download), and thus will not introduce new behaviors or change original behaviors. Additionally, the instrumentation locations depend on the fault types. For example, to manifest activity state loss, the $V$ operation will be added in \textsf{\small onStop} since state loss always happens after \textsf{\small onStop}.

\noindent\emph{\textbf{Example.}} The APE in \textsf{\small SearchByPartName\#onPostExecute} in Fig~\ref{fig:adsdroid} violates \emph{Rule 3}. To manifest this error, \tool instruments a $P$ operation in the end of \textsf{\small doInBackground}, and instruments a $V$ operation at the beginning of \textsf{\small SearchPanel\#onDestroy}. By doing this, \tool can easily manifest this error by a two-event sequence, \ie, click the ``Search" button and rotate the screen.

\vspace{-2mm}
\section{Implementation}
\label{sec:impl}
\tool is implemented in \textsf{\small Java} (5K LOC) and \textsf{\small Python} (1K LOC), and built on several existing tools to automatically manifest APEs.
It uses Soot~\cite{soot} to build call graph, statically detects APEs based on the three fault patterns, and generates program traces. When generating program traces, it extends IC3~\cite{Octeau15} to handle inner classes, fragments, and other public classes to reduce false negatives. It currently considers acyclic traces, and sets the maximum number of traces to 10 and maximum trace length to 20. It utilizes Gator~\cite{RountevY14} (the \textsf{\small GUIHierarchyPrinterClient} client in particular) to set up the mapping relations between user event handlers and UI elements, and converts the handlers (in the generated trace) to an event sequence (\eg, click a button, choose an item in the list). The mapped UI elements usually have unique IDs or texts, which enables \tool to interact with them.
It now supports \textsf{\small{back}}, \textsf{\small{rotate}}, \textsf{\small{Home}} (send the app to background), \textsf{\small{long-press-Home-and-back}}, \textsf{\small{Screen}} (screen on/off) to tweak lifecycle.  Apktool~\cite{apktool} instruments semaphore P/V operations into the UI thread and async threads to control thread scheduling. \textsf{\small{UIAutomator}}~\cite{uiautomator} is used to execute tests, and Android Debugging Bridge (\textsf{\small{adb}})~\cite{adb} monitors whether the app throws the intended 9 exception types.

%% file: evaluation.tex
\vspace{-2mm}
\section{Evaluation}
\label{sec:eval}

We applied \tool on 40 real-world Android apps, and compared it with three state-of-the-art GUI testing tools (Monkey, Sapienz, and Stoat), the Google official static analysis tool Lint, and the state-of-the-art data race detection tool, EventRacer, to measure its effectiveness.
We aim to answer these research questions.
\begin{itemize}[leftmargin=*]
	\item \emph{\textbf{RQ1:}} How effective is \tool for detecting APEs in Android apps? Can Lint detect them? 
	\item \emph{\textbf{RQ2:}} How effective is \tool against existing GUI testing techniques (Monkey, Sapienz, and Stoat) for APEs?
	\item \emph{\textbf{RQ3:}} How effective is \tool against existing data race detection techniques, for detecting specific types of APEs?
\end{itemize}

\vspace{-3mm}
\subsection{Evaluation Setup}

\begin{table*}[t]
	\footnotesize
	\newcommand{\tabincell}[2]{\begin{tabular}{@{}#1@{}}#2\end{tabular}}
	\renewcommand{\arraystretch}{1}
	
	\caption{Subjects used in the experiment and evaluation results of APEChecker and Lint.}
	\vspace{-3mm}
	\label{table:table_subjects}
	\centering
	\begin{tabular}{|c||c|c|c|c|c|c|c|c|c|c|c|}
		\hline
		App Name &\#ELOC &\tabincell{c}{\#Classes\\} &\#Methods &\#Activities & \tabincell{c}{\#Time\\(min)}  & \tabincell{c}{\#APEs\\(detected)}& \tabincell{c}{\#APEs\\(processed)} & \#Repro. &\#FP & \tabincell{c}{Fault\\Type} & \tabincell{c}{\#APEs\\ by Lint} \\
		\hline \hline
		\emph{Open Manager} &11477 & 56&227 &6 &1.0 & 2 & 2& 2&0 & 3 &0 \\
		\emph{ADSdroid} &2310  & 17 & 60 & 2&1.0 & 2 &2 & 2& 0& 3 &0\\
		\emph{DeskCon} & 11264& 80& 317& 5& 9.8&7 &5 & 4&0 & 3&0\\
		\emph{TuCanMobile} & 24264&113 &494 &10 & 3.6 &2 &2 &2 & 0 &3 & 0 \\
		\emph{RadioDroid} &19248  & 109 & 443 &2 & 2.8 & 3 & 1 &1 &2 &3 &0\\
		\emph{filmChecker} & 2739&28 &67 &2 &2.9 & 1& 1& 1&0 &3 &0\\
		\emph{QuranForMyAndroid} & 14170& 99&414 &11 &2.5 &4 & 4 &2 &0 &2, 3 &0\\
		\emph{MoTAC} &30857 &135 &748 &6 &1.8 &3& 2& 2&0 &2 &0\\
		\emph{MaximaOnAndroid} & 11357 & 43 &211 &4 & 2.5& 2& 2& 2 &0 &3&0 \\
		\emph{DebianDroid} & 14133 & 88 & 413 & 3& 3.1& 8 & 0 &0 &0 &-&0 \\
		\emph{NextGIS Mobile} & 12150 & 71 & 300 & 3& 3.2& 1&0 & 0& 0&-&0 \\
		\emph{A2DP Volume} &21502  &115  &555  &6 &3.7 &6  & 6&5 &0 &2, 3 &0\\
		\emph{Andor} &87818  & 445 & 2198 &19 & 3.0& 4 &3 & 3&1 &3&0 \\
		\emph{Mitzuli} &23615  &135  &517  &2 &3.4 & 2 & 1& 1 & 0&3&0 \\
		\emph{Commons} &32767 &177 &912 &9 &4.1 &3& 1& 1 & 0&3&0 \\
		\emph{AdAway} & 21443 & 131 & 535 &8 & 4.2& 2& 0& 0&0 &- &0\\
		\emph{ServeStream} & 64866 &300 &1777 & 8&4.8 &1& 1& 1& 0&3&0 \\
		\emph{Navit} &19336 &69 &360 &3 & 1.7 &2 & 0&0 &1 &- &0\\
		\emph{JKU App} & 52654 & 223 & 1263 &7 &5.2 & 2 & 0& 0 &0 &-&0 \\
		\emph{HomeManager} &9184  & 54 & 185 & 2 & 1.8 &  1& 1& 1& 0 & 1&0\\
		\emph{Transports Bordeaux} & 21840 & 180 & 668 &16 &7.3 & 3& 1 & 0&1 &-&0 \\
		\emph{MTG Familiar} &  84544& 341 & 1529 & 12&7.2 & 1 & 0& 0&0 &-&0 \\
		\emph{Cowsay} & 4149 &22  & 92& 1&1.1&1 & 1&1 & 0&1 &1\\
		\emph{AeonDroid} & 23037 &128  &604  &4 &3.6 &3& 1 & 1 &0 &3 &0\\
		\emph{Addi} &138716  &404  &2119 &2 &1.4 &1& 0 & 0& 0&- &0\\
		\hline
		\textbf{Average} &\textbf{30374}  &\textbf{143}  &\textbf{680} &\textbf{6} &\textbf{3.5} &\textbf{2.8}  &\textbf{-} &\textbf{-}& \textbf{-}&\textbf{-} &\textbf{-} \\
		\hline \hline
		\emph{MalayalamNewspaper} &7093  & 51 & 181 & 4&8.2 & 1& 1& 1&0 &3&0 \\
		\emph{Drum Solo} &17819 &80  &270  &4 &2.6 & 1 & 0&0 & 0& -&0\\
		\emph{Smart Poker} & 38835 & 219 & 1020 & 7&5.1 & 3& 3&1 & 0&3 &0\\
		\emph{Lojas Renner} & 24758&210 &756&11 &5.6 & 1 & 1& 1& 0 &3 &0\\
		\emph{Messaging} &114001 &446 &3103&13 &13.2 &1  &0 &0 & 0&- &0\\
		\emph{Recarga Vivo} &61913 &369 &1809&25 & 13.8&1& 0& 0&0 &-&0\\
		\emph{InstaCartoonPhoto} &1693 &26 & 60& 6&6.8 & 1& 1& 1&0 &3&0\\

		\emph{Fingerprint Lock} &17833 &123 &475&13 &6.9 &1& 0 &0 &0 &- &0\\
		\emph{Salmos} &19527& 110 &402 & 4 & 5.9&2& 2 & 2& 0&3 &0\\
		\emph{Santander} & 166101 &902  &3457  &5 &9.8 & 3& 2&1 &0 &3 &0 \\
		\emph{Biblia Sagrada} &41170&153  &648  &10 &16.6 &  7& 2& 2 & 1 &3 &0\\
		\emph{Trade Accounting} &85485  &450  &2018  &34 &11.8 & 2& 0& 0&0 &-&0\\
		\emph{PremiumWallpaper} & 52156 & 203 & 1321 & 10& 8.3& 4& 4& 4&0 &2, 3&0\\
        \emph{Tebak Lagu} & 23017 & 158 & 667 &12 & 11.5&9 & 7& 5& 0&2, 3 &0\\
        \emph{WorldNews Live24} & 104436 & 515 & 2298 &10 & 19.8& 3& 1& 1& 0& 3&0\\
		\hline
		\textbf{Average} &\textbf{51722}  &\textbf{267}  &\textbf{1232} &\textbf{12} &\textbf{9.9} &\textbf{2.7} &\textbf{-}  &\textbf{-}& \textbf{-}&\textbf{-} &\textbf{-}\\
		
		\hline
	\end{tabular}
	\vspace{-3mm}
\end{table*}

\noindent{\emph{\textbf{Subjects}}}.
We choose subjects from (1) F-droid,  the largest repository for open-source apps; and (2) Google Play Store, the official app store from Google.
We crawled all 2097 unique apps from F-droid, and 3107 popular apps from Google Play with over 10K installations.
Soot successfully processed 1654 F-droid apps and 2719 Google Play apps. Among them, 930 F-droid apps and 1274 Google Play apps use async constructs.
Next, \tool identifies 866 APEs in 234 F-droid apps (25.2\%=234/930) and 1161 APEs in 201 Google Play apps (15.8\%=201/1274) that contain suspicious APEs, respectively.

\noindent{\emph{\textbf{Environment}}}.
\tool runs on a 64-bit Ubuntu 14.04 machine
with 12 cores (3.50GHz Intel CPU) and 32GB RAM.
We verify APEs on both an Android emulator (SDK 4.4.2) and an LG Nexus 5X mobile phone (SDK 7.1.1).

\noindent{\emph{\textbf{Studies}}}. We conducted three case studies.
In \emph{\textbf{Study 1}}, we answer \emph{\textbf{RQ1}}. 
We need to manually analyze each reported APE, and determine the true positives and false positives, which requires a lot of human efforts. Therefore, we randomly selected around 10\% F-droid and Google Play apps, respectively (the apps requiring user credentials are excluded): (1) 25 F-droid apps from 234 suspicious faulty apps, with 30374 executable lines of code (LOC) (in \textsf{\small{Jimple}}), 143 classes, 680 methods and 6 activities on average; and (2) 15 Google Play apps with 51722 LOC, 267 classes, 1232 methods and 12 activities on average, shown in Table~\ref{table:table_subjects}. We measured the static analysis time (the time of replay tests is omitted, since it only takes a few seconds), the number of detected APEs, the number of APEs that can be processed (limited by the abilities of Soot, IC3, Gator), the number of APEs that can be reproduced (\#Repro.), false positives (\#FP), the types of faults, and the number of APEs that can be detected by Lint.

In \emph{\textbf{Study 2}}, we answer \emph{\textbf{RQ2}} by comparing \tool with Monkey, Sapienz, and Stoat, which have proven effectiveness on bug detection~\cite{ChoudharyGO15,MaoHJ16,Stoat2017}.
10 apps are randomly selected from 234 F-droid apps that contain APEs. Each tool is allocated with one hour with default settings.
The number of confirmed (\ie, successfully triggered) APEs, the analysis time and the length of tests (\ie, event sequences) are recorded.
From the perspective of their approach workflow, all tools are given the original rather than the instrumented apps (\cf Section~\ref{sec:errorv}) as input to achieve fair comparison --- Monkey, Sapienz and Stoat do not require the instrumentation for bug detection as their algorithms do not need instrumentation (otherwise they may be adversely affected, \eg, stuck by the thread scheduling), and also do not have idea of where to instrument.
To alleviate the randomness, we run each tool 10 times to average data.

In \emph{\textbf{Study 3}}, we answer \emph{\textbf{RQ3}} by comparing \tool with EventRacer~\cite{Bielik15}. We did not choose other data race tools, \eg, CAFA~\cite{HsiaoPYPNCKF14} (not available) and DroidRacer~\cite{MaiyaKM14,droidracer} (less effecitve and precise than EventRacer~\cite{Bielik15}).
In detail, we (1) randomly selected 10 F-droid apps  (since the source code is required to confirm the APEs reported by EventRacer, listed in Table~\ref{table:eventracer}) with at least one APE of \emph{Fault Pattern 3} (treated as data races by EventRacer); (2) only consider the races reported between UI thread and async threads;
and (3) allocated 10,000 events for EventRacer (by default, only 1,000 events) to generate event sequences, which on average costs 15 minutes per app (comparable to the running time of \tool). EventRacer generates a lot of false positives, but to our knowledge, no reproducing tools can facilitate our analysis: AsyncDroid~\cite{OzkanET15} requires user-provided event traces, RacerDroid~\cite{Tang2016} adopts manually analysis and ERVA~\cite{Hu2016} is not available.
Therefore, we manually analyzed the races, and resorted to developers for confirmation.

\vspace{-3mm}
\subsection{Study 1: Effectiveness of Detecting APEs}
\label{sec:study1}
Table~\ref{table:table_subjects} shows the results of \tool on the 40 apps. It identified 107 APEs in total with 67 and 40 from F-droid and Google Play apps, respectively. 
Among them, due to the limitations of underlying tools, 61 APEs can be processed, of which 51 are successfully confirmed with real tests, achieving 83.6\% (51/61) hit rate.
On average, it takes 6.7 minutes for one app, with 3.5 minutes for one F-droid app and 9.9 minutes for one Google Play app. In contrast, Lint only detects one APE, which is ineffective.
We detail the results below.

\noindent{\emph{\textbf{Effectiveness.}}}
\tool successfully reproduces all APEs of \emph{Fault Pattern 1}.
\textsf{\small HomeManager} manages and switches between home applications. Once started, it scans the installed apps on the device in an async thread, which modifies the data list of app info asychronously (with a \textsf{\small ListView}). When the scan is finished, it notifies the UI thread to refresh the \textsf{\small ListView} with the new list. However, if the list is changed again (another async thread is created) during the refresh, the app can be crashed by an \textsf{\small IllegalStateException}. 

In \emph{Fault Pattern 2}, \emph{MTG Familiar} is an offline database app for magic cards. It takes the photos of cards to identify the magic card. After taking a photo, it starts loading the photo with a \textsf{\small ProgressDialog} and calls \textsf{\small Toast\#makeText} in an async thread. However, it fails to post \textsf{\small Toast} creation onto the UI thread, which causes a \textsf{\small RuntimeException}.

\tool successfully reproduces most APEs of \emph{Fault Pattern 3}. For example, 
\emph{Mitzuli}, a translator app, starts an async thread to check version update, and shows a dialog to notify the results.
However, if users rotate the device before the thread returns, 
a fatal \textsf{\small BadTokenException} occurs.

\noindent{\emph{\textbf{Hit Rate.}}}
\tool achieves an 83.6\% (51/61) hit rate.
We analyzed the failed cases, and found most of them require special constraints.
We summarize the main scenarios below.

\noindent \emph{(1) External environment.} 
\emph{Transports Bordeaux}, a transportation app, searches for paths from the starts to destinations. \tool can generate a program trace but cannot generate an event sequence to trigger the error, since the app registers a \textsf{\small Broadcast Receiver} for receiving specific intents. \tool  cannot infer this now. 
\emph{QuranForMyAndroid} is a translator for ``The Quran". If ``Back" button is pressed after editting the text, it starts a thread to save the text into storage and shows a message via \textsf{\small Toast\#makeText} only when no data storage is available. \tool cannot reproduce such errors. 

\noindent \emph{(2) Complicated Exceptions.} 
\tool currently cannot infer complicated exceptions such as file not found and database corruption. One APE in \emph{A2DP Volume} is that it tries to show a message via \textsf{\small Toast\#makeText} in the exception handling's \textsf{\small catch} statement which, however, is only reachable when the database to be loaded is corrupted in \textsf{\small try} statement.

\noindent{\emph{\textbf{False Positives.}}}
\tool only reports 6 false positives (5.6\%=6/107). The main reason is that some methods or classes are actually unreachable. For example, in \emph{RadioDroid}, \tool identifies 3 potential APEs, 2 of which are false positive since they reside in a dead activity (\textsf{\small ActivityRadioStationDetail}) that cannot be started by any other activities.

\noindent\emph{In summary, on the evaluated apps, \tool can efficiently manifest APEs in a few minutes with 83.6\% hit rate and 5.6\% false alarms.}

\begin{figure}[t]
	\begin{center}
		\includegraphics[width=0.45\textwidth]{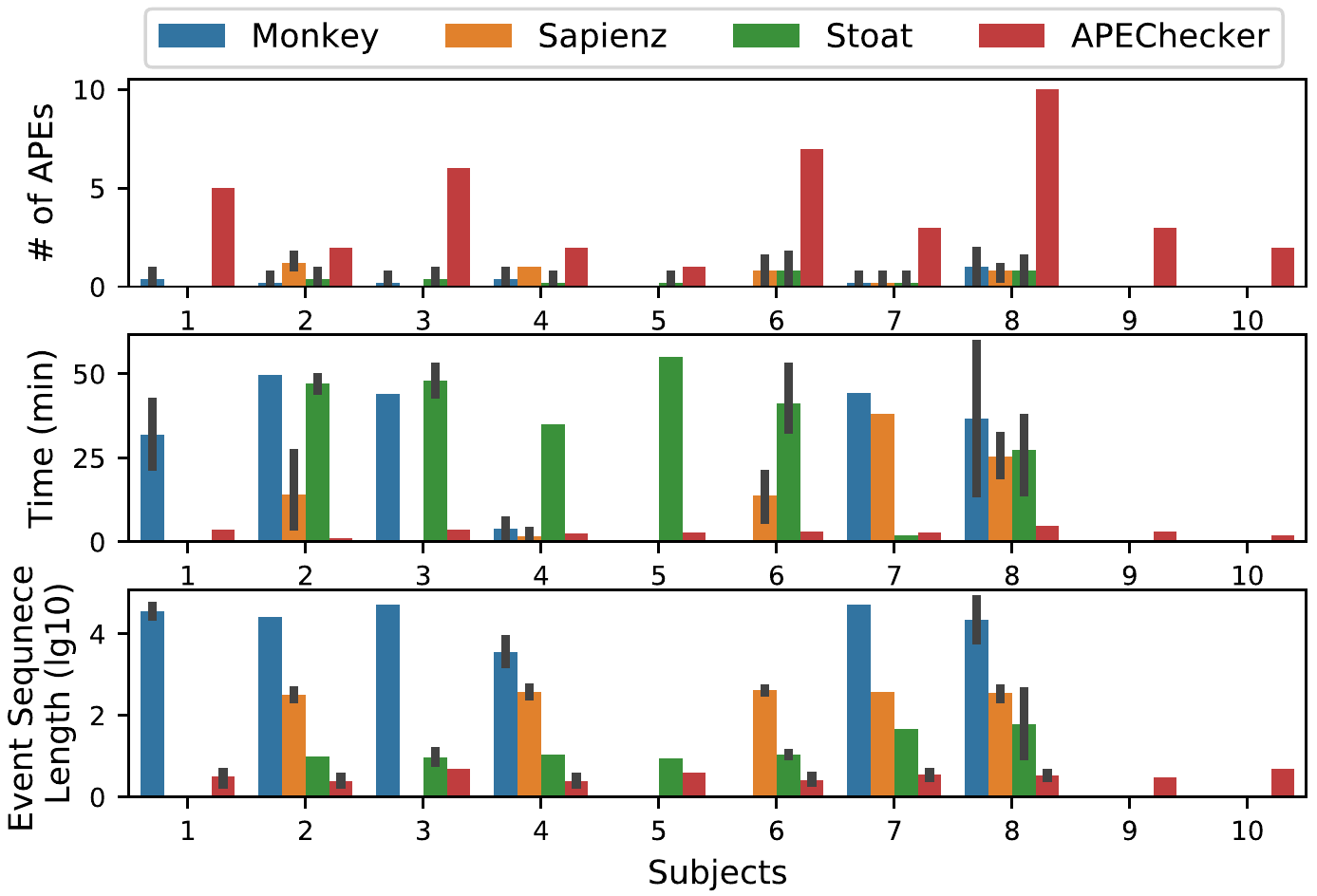}
	\end{center}
	\vspace{-4mm}
	\caption{Comparison between \tool and testing tools.}
	\label{fig:testing_comparison}
	\vspace{-6mm}
\end{figure}

\vspace{-2mm}
\subsection{Study 2: Comparison with Testing Tools}

Fig.~\ref{fig:testing_comparison} compares \tool and the three testing tools in the number of confirmed APEs, the analysis time (in minutes) and the length of event sequence (in logarithmic scale) that triggers the errors. The results are presented in three bar plots with error bars (shown as black lines), where each bar's height indicates the average value, and the error bar represents the standard deviation of uncertainty. 

\tool confirms many more APEs than the testing tools. In the 10 runs, \tool totally confirms 32 unique APEs from these 10 apps, while Monkey, Sapienz and Stoat, respectively, confirm 7, 8, and 9.
Moreover, none of the testing tools uncover APEs in subjects \emph{9} and \emph{10} while \tool finds 3 and 2 ones, respectively. The error bars of testing tools show obvious data variance, \ie, testing tools are very likely to miss APEs.

For the testing tools, the analysis time of an error is its first occurrence time since one error can be triggered multiple times during one run. 
We can see \tool takes much less time than the testing tools except for the subjects \emph{4} and \emph{7}. On average, \tool takes 2.9 minutes, while Monkey, Sapienz, and Stoat require 32.7, 14.3 and 37.1 minutes, respectively.
Sapienz can easily hit one of the errors in the subject \emph{4} but take more time for the other twos. Stoat takes less time on the subject \emph{7}, but only hits that APE once during the 10 runs. Additionally, in terms of the event sequence length, \tool provides shorter (more usable) tests than the testing tools. The average event sequence length of Monkey, Sapienz, Stoat and \tool is 35725, 411, 12, 4, respectively.

\noindent\emph{In summary, on the evaluated apps, \tool detects 3X more APEs than testing tools, reduces detection time from half an hour to a few minutes, and provides more usable reproducing tests.}

\vspace{-2mm}
\subsection{Study 3: Comparison with EventRacer}
Table~\ref{table:eventracer} compares \tool and EventRacer on 10 F-droid apps in the number of detected APEs. For EventRacer, we give the number of reported data races and true positives under two configurations (1K-event and 10K-event).
To confirm true positives, we followed the approach of EVRA~\cite{Hu2016} and remove duplicated races: if data races are reported on different variables in ADF but refer to the same location of app code, we treat them as one true positive. Because the races actually indicate the same error in the app. 

From Table~\ref{table:eventracer}, we can see \tool is more effective than EventRacer for the specific APEs of \emph{Fault Pattern 3}. \tool detects and reproduces 32 out of 38 APEs while EventRacer (10K-event) only finds 6 ones.
By comparing the results between 1K-event and 10K-event, we can note the effectiveness of EventRacer heavily depends on the number of given events. 
Although EventRacer can detect 5 more APEs with 10K-event, it reports 14X more false positives, which overwhelms users. This is also observed by EVRA~\cite{Hu2016} that only 3\% data races reported by EventRacer are harmful.
We also observe that the reported races are highly random, which further undermines its effectiveness for detecting these specfic APEs.

\noindent\emph{In summary, on the evaluated apps, \tool detects 5X more APEs than EventRacer on a specific APE type with very few false alarms.}

\begin{table}[t]
	\scriptsize
	\newcommand{\tabincell}[2]{\begin{tabular}{@{}#1@{}}#2\end{tabular}}
	\renewcommand{\arraystretch}{1}
	\caption{Comparison between \tool and EventRacer}
	\vspace{-3mm}
	\label{table:eventracer}
	\centering
	\begin{tabular}{|c|c|c|c|c|c|c|}
		\hline

\multicolumn{1}{|c|}{\multirow{2}{*}{App Name}} &\multicolumn{1}{c|}{\multirow{2}{*}{\begin{tabular}[c]{@{}c@{}}\#APEs\\(Patt.3)\end{tabular}}}&\multicolumn{1}{c|}{\multirow{2}{*}{\begin{tabular}[c]{@{}c@{}}APECh.\\\#Repro.\end{tabular}}}& \multicolumn{2}{c|}{EventRacer (1K)} & \multicolumn{2}{c|}{EventRacer (10K)}  \\ \cline{4-7} 
\multicolumn{1}{|c|}{} & \multicolumn{1}{c|}{} & \multicolumn{1}{c|}{} & \multicolumn{1}{c|}{\#Reported} & \multicolumn{1}{c|}{\#TP } & \multicolumn{1}{c|}{\#Reported} & \multicolumn{1}{c|}{\#TP } \\

		\hline \hline
		\emph{A2DPVolume} &5&5 & 41 &0&1249 &1 \\
		\emph{TuCanMobile} &2&2&12 &0 & 53&0\\
		\emph{OpenManager} &2&2 & 0&0 & 27&1\\
		\emph{Andor} &4&3&0 &0 & 0&0\\
		\emph{Maxima} &2 &2& 1&0&36 &1 \\
		\emph{filmChecker} &1&1&36 &0&139 &0 \\
		\emph{DeskCon} &7&5& 0&0&29 &1 \\
		\emph{RadioDroid} &3&1&15 &1 &48 &1\\
		\emph{ADSdroid} &2&2&2 &0&1 &0\\
		\emph{AnyMemo} & 10&9& 19&0 &206 &1\\
		
		\hline
		\textbf{Total} &\textbf{38}  &\textbf{32}  &\textbf{126} &\textbf{1} &\textbf{1788} &\textbf{6} \\

		\hline

	\end{tabular}
	\vspace{-4mm}
\end{table}
\vspace{-2mm}
\subsection{Limitations and Threats to Validity}
\noindent\textbf{Limitations.} \tool builds on several popular research tools, \ie, Soot, IC3, and Gator, which represent the state-of-the-art. However, they still have several limitations in practice, which directly fail \tool on program trace generation. For example, even if Soot successfully processes an app,  the call graph can still be incomplete (21.3\% cases in our evaluation). Moreover, IC3 cannot handle ICCs in inner classes or fragments, and Gator cannot handle third-party UI controls. To counter this, we have already extended IC3 and Gator in these aspects (Section~\ref{sec:impl}). But \tool still cannot generate valid program traces for 46 APEs detected in the evaluated apps (\eg, \emph{DebianDroid}) due to the limitations of underlying tools. We believe more engineering efforts on these tools can significantly improve \tool.
Additionally, \tool now only infers simple environment, which lowers the hit rate. Future work will integrate more sophisticated analysis techniques to overcome this. Aditionally, applying \tool to detect APEs for other GUI frameworks like \textsf{\small Qt} requires proper extensions~\cite{SanthiarKK16}.

\noindent\textbf{Threats to Validity.} 
\tool is only evaluated on 40 apps, and thus our conclusion may not be general to all apps. The summarized async programming rules and fault patterns may be incomplete, and thus \tool may suffer from false negatives. But we have thoroughly inspected all relevant resources, and the fault patterns have covered all 375 real APEs from 1091 apps. To ease the verification of APEs, \tool simulates specific execution environment by controlling thread scheduling (\eg, increase the wait time of network access). So some triggered APEs may not be easily manifested by users although they are real faults under specific conditions.

%% file: relwork.tex
\vspace{-2mm}
\section{Related Work}
\label{sec:rel}

\noindent{\emph{\textbf{Automated GUI Testing}}} Many GUI testing techniques, \eg, random testing~\cite{MachiryTN13}, search-based testing~\cite{MahmoodMM14,MaoHJ16}, symbolic execution~\cite{AnandNHY12,vanderMerwe2012,MirzaeiMPEM12}, model-based testing~\cite{AmalfitanoFTCM12,YangPX13,AzimN13,ChoiNS13,AmalfitanoFTTM15,FSMdroid16,Stoat2017} and other approaches~\cite{VasquezWBMP15,Mirzaei16,Song2017,MaoHJ17}, have been proposed for Android apps. As we discussed in Section~\ref{sec:intro} and demonstrated in the evaluation, these techniques are ineffective for APEs due to lack of prior knowledge (fault patterns). In contrast, \tool leverages this knowledge to efficiently manifest APEs.

\tool tackles specific bugs that violate the rules of single-GUI-thread model. 
QUANTUM~\cite{Zaeem14} uses model-based testing to tackle app-agnostic bugs. For example, when rotation happens, the screen should show the same content and support the actions as before.
Thor~\cite{Adamsen15} amplifies existing tests by randomly injecting neutral event sequences (\eg, double rotation, pause-and-resume) that should not affect the outputs of original tests. 
Amalfitano \etal~\cite{AmalfitanoRPF18} propose a similar idea by exhaustively injecting orientation changes to expose GUI failures.
However, these techniques may not be effective for APEs.
First, their effectiveness depends on the quality of an existing GUI model or test suite, which are usually not available~\cite{KochharTNZL15}. Second, these techniques are random, and do not explicitly consider the interactions between UI and aysnc threads.
Graziussi~\cite{lifecycle16} encodes some specific patterns into Lint to detect lifecycle bugs.
KREfinder~\cite{Shan16} finds restart-and-resume errors that cause app data loss. 
\tool tackles APEs which have not been systematically investigated before.

\noindent\emph{\textbf{Concurreny Bugs}}
As discussed in Section~\ref{sec: prior_work}, some APEs can be treated as concurrency bugs.
To tackle them, a number of data race detection tools~\cite{HsiaoPYPNCKF14,MaiyaKM14,Bielik15,Safi15} and data race reproducing tools~\cite{OzkanET15,Tang2016,Hu2016} are developed.
However, as our evaluation shows, data race detection tools are ineffective for these APEs.
AsyncDroid~\cite{OzkanET15} and RacerDroid~\cite{Tang2016} are impractical since they either require user-provided traces or depend on
the results of data race detection tools.
ERVA~\cite{Hu2016} utilizes dependency graph and event flipping to reproduce races, and uses state comparison techniques to identify harmful ones. 
AATT~\cite{Li0GXMML16} manifests concurrency bugs whose race points at the app code level. However, it can only manifest races whose conflicting events are in the same page or race points in the user code. 
Other bug reproducing tools like RERAN~\cite{GomezNAM13} and CrashScope~\cite{MoranVBVP16} require user interactions.
In contrast, \tool automatically generates event sequences to verify APEs.

\noindent\emph{\textbf{Guided Test Generation}}
Our work uses static analysis to locate suspicious APEs, and then guides GUI interactions to verify them.
This idea is also adopted in other application domains of Android.
SmartDroid~\cite{Zheng12} guides tests to reach certain APIs to manifest the malicious behaviors. However, it explores every UI element on each page to find the right view to click. Brahmastra~\cite{Bhoraskar14} drives apps to reach third-party components to test security issues; MAMBA~\cite{Keng2016} guides tests to reveal potential accesses to privacy-sensitive data;
FuzzDroid~\cite{Rasthofer2017} utilizes a search-based algorithm combined with static and dynamic analysis to manifest malicious behaviors.
However, none of these tools can be directly compared with \tool for detecting APEs.
Other targeted testing tools include Collider~\cite{JensenPM13}, ConDroid~\cite{SchutteFT15}, A$^3$E~\cite{AzimN13}.
However, they are either not available or requiring human intervention.

%% file: conclusion.tex
\vspace{-2.5mm}
\section{Conclusion}
\label{sec:concl}
This paper introduces \tool, a technique to efficiently manifest APEs.
First, we conduct a formative study to understand the aysnc programming rules implied by the single-GUI-thread model, and collect a set of real APE issues. 
Second, informed by these results, we distill three fault patterns, and locate suspicious APEs via static fault pattern analysis, and then verify them with real tests. The evaluation shows \tool is useful and effective.

%% file: sample-sigconf.bbl

\begin{thebibliography}{89}


\ifx \showCODEN    \undefined \def \showCODEN     #1{\unskip}     \fi
\ifx \showDOI      \undefined \def \showDOI       #1{#1}\fi
\ifx \showISBNx    \undefined \def \showISBNx     #1{\unskip}     \fi
\ifx \showISBNxiii \undefined \def \showISBNxiii  #1{\unskip}     \fi
\ifx \showISSN     \undefined \def \showISSN      #1{\unskip}     \fi
\ifx \showLCCN     \undefined \def \showLCCN      #1{\unskip}     \fi
\ifx \shownote     \undefined \def \shownote      #1{#1}          \fi
\ifx \showarticletitle \undefined \def \showarticletitle #1{#1}   \fi
\ifx \showURL      \undefined \def \showURL       {\relax}        \fi
\providecommand\bibfield[2]{#2}
\providecommand\bibinfo[2]{#2}
\providecommand\natexlab[1]{#1}
\providecommand\showeprint[2][]{arXiv:#2}

\bibitem[\protect\citeauthoryear{Adamsen, Mezzetti, and M{\o}ller}{Adamsen
  et~al\mbox{.}}{2015}]%
        {Adamsen15}
\bibfield{author}{\bibinfo{person}{Christoffer~Quist Adamsen},
  \bibinfo{person}{Gianluca Mezzetti}, {and} \bibinfo{person}{Anders
  M{\o}ller}.} \bibinfo{year}{2015}\natexlab{}.
\newblock \showarticletitle{Systematic Execution of Android Test Suites in
  Adverse Conditions}. In \bibinfo{booktitle}{\emph{Proceedings of the 2015
  International Symposium on Software Testing and Analysis}}
  \emph{(\bibinfo{series}{ISSTA 2015})}. \bibinfo{publisher}{ACM},
  \bibinfo{address}{New York, NY, USA}, \bibinfo{pages}{83--93}.
\newblock
\showISBNx{978-1-4503-3620-8}


\bibitem[\protect\citeauthoryear{Amalfitano, Fasolino, Tramontana, Carmine, and
  Memon}{Amalfitano et~al\mbox{.}}{2012}]%
        {AmalfitanoFTCM12}
\bibfield{author}{\bibinfo{person}{Domenico Amalfitano},
  \bibinfo{person}{Anna~Rita Fasolino}, \bibinfo{person}{Porfirio Tramontana},
  \bibinfo{person}{Salvatore~De Carmine}, {and} \bibinfo{person}{Atif~M.
  Memon}.} \bibinfo{year}{2012}\natexlab{}.
\newblock \showarticletitle{Using {GUI} ripping for automated testing of
  {Android} applications}. In \bibinfo{booktitle}{\emph{{IEEE/ACM}
  International Conference on Automated Software Engineering, ASE'12, Essen,
  Germany, September 3-7, 2012}}. \bibinfo{pages}{258--261}.
\newblock


\bibitem[\protect\citeauthoryear{Amalfitano, Fasolino, Tramontana, Ta, and
  Memon}{Amalfitano et~al\mbox{.}}{2015}]%
        {AmalfitanoFTTM15}
\bibfield{author}{\bibinfo{person}{Domenico Amalfitano},
  \bibinfo{person}{Anna~Rita Fasolino}, \bibinfo{person}{Porfirio Tramontana},
  \bibinfo{person}{Bryan~Dzung Ta}, {and} \bibinfo{person}{Atif~M. Memon}.}
  \bibinfo{year}{2015}\natexlab{}.
\newblock \showarticletitle{{MobiGUITAR}: Automated Model-Based Testing of
  Mobile Apps}.
\newblock \bibinfo{journal}{\emph{{IEEE} Software}} \bibinfo{volume}{32},
  \bibinfo{number}{5} (\bibinfo{year}{2015}), \bibinfo{pages}{53--59}.
\newblock


\bibitem[\protect\citeauthoryear{Amalfitano, Riccio, Paiva, and
  Fasolino}{Amalfitano et~al\mbox{.}}{2018}]%
        {AmalfitanoRPF18}
\bibfield{author}{\bibinfo{person}{Domenico Amalfitano},
  \bibinfo{person}{Vincenzo Riccio}, \bibinfo{person}{Ana C.~R. Paiva}, {and}
  \bibinfo{person}{Anna~Rita Fasolino}.} \bibinfo{year}{2018}\natexlab{}.
\newblock \showarticletitle{Why does the orientation change mess up my Android
  application? From {GUI} failures to code faults}.
\newblock \bibinfo{journal}{\emph{Softw. Test., Verif. Reliab.}}
  \bibinfo{volume}{28}, \bibinfo{number}{1} (\bibinfo{year}{2018}).
\newblock


\bibitem[\protect\citeauthoryear{Anand, Naik, Harrold, and Yang}{Anand
  et~al\mbox{.}}{2012}]%
        {AnandNHY12}
\bibfield{author}{\bibinfo{person}{Saswat Anand}, \bibinfo{person}{Mayur Naik},
  \bibinfo{person}{Mary~Jean Harrold}, {and} \bibinfo{person}{Hongseok Yang}.}
  \bibinfo{year}{2012}\natexlab{}.
\newblock \showarticletitle{Automated concolic testing of smartphone apps}. In
  \bibinfo{booktitle}{\emph{20th {ACM} {SIGSOFT} Symposium on the Foundations
  of Software Engineering (FSE-20), SIGSOFT/FSE'12, Cary, NC, {USA} - November
  11 - 16, 2012}}. \bibinfo{pages}{59}.
\newblock


\bibitem[\protect\citeauthoryear{Apple}{Apple}{2018a}]%
        {ios}
\bibfield{author}{\bibinfo{person}{Apple}.} \bibinfo{year}{2018}\natexlab{a}.
\newblock \bibinfo{title}{{App Programming Guide for iOS}}.
\newblock
\newblock
\urldef\tempurl%
\url{https://developer.apple.com/library/content/navigation/}
\showURL{%
Retrieved 2018-7 from \tempurl}


\bibitem[\protect\citeauthoryear{Apple}{Apple}{2018b}]%
        {macos}
\bibfield{author}{\bibinfo{person}{Apple}.} \bibinfo{year}{2018}\natexlab{b}.
\newblock \bibinfo{title}{{MacOS Cocoa}}.
\newblock
\newblock
\urldef\tempurl%
\url{https://developer.apple.com/macos/}
\showURL{%
Retrieved 2018-7 from \tempurl}


\bibitem[\protect\citeauthoryear{Azim and Neamtiu}{Azim and Neamtiu}{2013}]%
        {AzimN13}
\bibfield{author}{\bibinfo{person}{Tanzirul Azim} {and} \bibinfo{person}{Iulian
  Neamtiu}.} \bibinfo{year}{2013}\natexlab{}.
\newblock \showarticletitle{Targeted and depth-first exploration for systematic
  testing of {Android} apps}. In \bibinfo{booktitle}{\emph{Proceedings of the
  2013 {ACM} {SIGPLAN} International Conference on Object Oriented Programming
  Systems Languages {\&} Applications, {OOPSLA} 2013, part of {SPLASH} 2013,
  Indianapolis, IN, USA, October 26-31, 2013}}. \bibinfo{pages}{641--660}.
\newblock


\bibitem[\protect\citeauthoryear{Bhoraskar, Han, Jeon, Azim, Chen, Jung, Nath,
  Wang, and Wetherall}{Bhoraskar et~al\mbox{.}}{2014}]%
        {Bhoraskar14}
\bibfield{author}{\bibinfo{person}{Ravi Bhoraskar}, \bibinfo{person}{Seungyeop
  Han}, \bibinfo{person}{Jinseong Jeon}, \bibinfo{person}{Tanzirul Azim},
  \bibinfo{person}{Shuo Chen}, \bibinfo{person}{Jaeyeon Jung},
  \bibinfo{person}{Suman Nath}, \bibinfo{person}{Rui Wang}, {and}
  \bibinfo{person}{David Wetherall}.} \bibinfo{year}{2014}\natexlab{}.
\newblock \showarticletitle{Brahmastra: Driving Apps to Test the Security of
  Third-party Components}. In \bibinfo{booktitle}{\emph{Proceedings of the 23rd
  USENIX Conference on Security Symposium}} \emph{(\bibinfo{series}{SEC'14})}.
  \bibinfo{publisher}{USENIX Association}, \bibinfo{address}{Berkeley, CA,
  USA}, \bibinfo{pages}{1021--1036}.
\newblock
\showISBNx{978-1-931971-15-7}


\bibitem[\protect\citeauthoryear{Bielik, Raychev, and Vechev}{Bielik
  et~al\mbox{.}}{2015}]%
        {Bielik15}
\bibfield{author}{\bibinfo{person}{Pavol Bielik}, \bibinfo{person}{Veselin
  Raychev}, {and} \bibinfo{person}{Martin Vechev}.}
  \bibinfo{year}{2015}\natexlab{}.
\newblock \showarticletitle{Scalable Race Detection for Android Applications}.
  In \bibinfo{booktitle}{\emph{Proceedings of the 2015 ACM SIGPLAN
  International Conference on Object-Oriented Programming, Systems, Languages,
  and Applications}} \emph{(\bibinfo{series}{OOPSLA 2015})}.
  \bibinfo{publisher}{ACM}, \bibinfo{address}{New York, NY, USA},
  \bibinfo{pages}{332--348}.
\newblock
\showISBNx{978-1-4503-3689-5}


\bibitem[\protect\citeauthoryear{Choi, Necula, and Sen}{Choi
  et~al\mbox{.}}{2013}]%
        {ChoiNS13}
\bibfield{author}{\bibinfo{person}{Wontae Choi}, \bibinfo{person}{George~C.
  Necula}, {and} \bibinfo{person}{Koushik Sen}.}
  \bibinfo{year}{2013}\natexlab{}.
\newblock \showarticletitle{Guided {GUI} testing of {Android} apps with minimal
  restart and approximate learning}. In \bibinfo{booktitle}{\emph{Proceedings
  of the 2013 {ACM} {SIGPLAN} International Conference on Object Oriented
  Programming Systems Languages {\&} Applications, {OOPSLA} 2013, part of
  {SPLASH} 2013, Indianapolis, IN, USA, October 26-31, 2013}}.
  \bibinfo{pages}{623--640}.
\newblock


\bibitem[\protect\citeauthoryear{Choudhary, Gorla, and Orso}{Choudhary
  et~al\mbox{.}}{2015}]%
        {ChoudharyGO15}
\bibfield{author}{\bibinfo{person}{Shauvik~Roy Choudhary},
  \bibinfo{person}{Alessandra Gorla}, {and} \bibinfo{person}{Alessandro Orso}.}
  \bibinfo{year}{2015}\natexlab{}.
\newblock \showarticletitle{Automated Test Input Generation for {Android}: Are
  We There Yet? {(E)}}. In \bibinfo{booktitle}{\emph{30th {IEEE/ACM}
  International Conference on Automated Software Engineering, {ASE} 2015,
  Lincoln, NE, USA, November 9-13, 2015}}. \bibinfo{pages}{429--440}.
\newblock


\bibitem[\protect\citeauthoryear{CodePath}{CodePath}{2018}]%
        {codepath}
\bibfield{author}{\bibinfo{person}{CodePath}.} \bibinfo{year}{2018}\natexlab{}.
\newblock \bibinfo{title}{{CodePath Android Cliffnotes}}.
\newblock
\newblock
\urldef\tempurl%
\url{http://guides.codepath.com/android}
\showURL{%
Retrieved 2018-7 from \tempurl}


\bibitem[\protect\citeauthoryear{developers}{developers}{2018}]%
        {apktool}
\bibfield{author}{\bibinfo{person}{Apktool developers}.}
  \bibinfo{year}{2018}\natexlab{}.
\newblock \bibinfo{title}{{A tool for reverse engineering Android apk files}}.
\newblock
\newblock
\urldef\tempurl%
\url{https://ibotpeaches.github.io/Apktool/}
\showURL{%
Retrieved 2018-7 from \tempurl}


\bibitem[\protect\citeauthoryear{Developers}{Developers}{2018a}]%
        {adsdroid}
\bibfield{author}{\bibinfo{person}{Adsdroid Developers}.}
  \bibinfo{year}{2018}\natexlab{a}.
\newblock \bibinfo{title}{{Adsdroid}}.
\newblock
\newblock
\urldef\tempurl%
\url{https://github.com/dnet/adsdroid}
\showURL{%
Retrieved 2018-7 from \tempurl}


\bibitem[\protect\citeauthoryear{Developers}{Developers}{2018b}]%
        {gisapp}
\bibfield{author}{\bibinfo{person}{Android~Gisapp Developers}.}
  \bibinfo{year}{2018}\natexlab{b}.
\newblock \bibinfo{title}{{Android Gisapp}}.
\newblock
\newblock
\urldef\tempurl%
\url{https://github.com/nextgis/android_gisapp}
\showURL{%
Retrieved 2018-7 from \tempurl}


\bibitem[\protect\citeauthoryear{Developers}{Developers}{2018c}]%
        {pedometer}
\bibfield{author}{\bibinfo{person}{Privacy Friendly~Pedometer Developers}.}
  \bibinfo{year}{2018}\natexlab{c}.
\newblock \bibinfo{title}{{Privacy Friendly Pedometer}}.
\newblock
\newblock
\urldef\tempurl%
\url{https://github.com/SecUSo/privacy-friendly-pedometer}
\showURL{%
Retrieved 2018-7 from \tempurl}


\bibitem[\protect\citeauthoryear{Fan, Su, Chen, Meng, Liu, Xu, Pu, and Su}{Fan
  et~al\mbox{.}}{2018}]%
        {fan18}
\bibfield{author}{\bibinfo{person}{Lingling Fan}, \bibinfo{person}{Ting Su},
  \bibinfo{person}{Sen Chen}, \bibinfo{person}{Guozhu Meng},
  \bibinfo{person}{Yang Liu}, \bibinfo{person}{Lihua Xu},
  \bibinfo{person}{Geguang Pu}, {and} \bibinfo{person}{Zhendong Su}.}
  \bibinfo{year}{2018}\natexlab{}.
\newblock \showarticletitle{Large-scale analysis of framework-specific
  exceptions in Android apps}. In \bibinfo{booktitle}{\emph{Proceedings of the
  40th International Conference on Software Engineering, {ICSE} 2018,
  Gothenburg, Sweden, May 27 - June 03, 2018}}. \bibinfo{pages}{408--419}.
\newblock


\bibitem[\protect\citeauthoryear{FindBugs}{FindBugs}{2018}]%
        {findbugs}
\bibfield{author}{\bibinfo{person}{FindBugs}.} \bibinfo{year}{2018}\natexlab{}.
\newblock \bibinfo{title}{{FindBugs}}.
\newblock
\newblock
\urldef\tempurl%
\url{http://findbugs.sourceforge.net/}
\showURL{%
Retrieved 2018-7 from \tempurl}


\bibitem[\protect\citeauthoryear{Gomez, Neamtiu, Azim, and Millstein}{Gomez
  et~al\mbox{.}}{2013}]%
        {GomezNAM13}
\bibfield{author}{\bibinfo{person}{Lorenzo Gomez}, \bibinfo{person}{Iulian
  Neamtiu}, \bibinfo{person}{Tanzirul Azim}, {and} \bibinfo{person}{Todd~D.
  Millstein}.} \bibinfo{year}{2013}\natexlab{}.
\newblock \showarticletitle{{RERAN:} timing- and touch-sensitive record and
  replay for Android}. In \bibinfo{booktitle}{\emph{35th International
  Conference on Software Engineering, {ICSE} '13, San Francisco, CA, USA, May
  18-26, 2013}}. \bibinfo{pages}{72--81}.
\newblock


\bibitem[\protect\citeauthoryear{Google}{Google}{2018a}]%
        {activity}
\bibfield{author}{\bibinfo{person}{Google}.} \bibinfo{year}{2018}\natexlab{a}.
\newblock \bibinfo{title}{{Activity Lifecycle}}.
\newblock
\newblock
\urldef\tempurl%
\url{https://developer.android.com/guide/components/activities/activity-lifecycle.html}
\showURL{%
Retrieved 2018-7 from \tempurl}


\bibitem[\protect\citeauthoryear{Google}{Google}{2018b}]%
        {adb}
\bibfield{author}{\bibinfo{person}{Google}.} \bibinfo{year}{2018}\natexlab{b}.
\newblock \bibinfo{title}{{Android Debug Bridge}}.
\newblock
\newblock
\urldef\tempurl%
\url{https://developer.android.com/studio/command-line/adb.html}
\showURL{%
Retrieved 2018-7 from \tempurl}


\bibitem[\protect\citeauthoryear{Google}{Google}{2018c}]%
        {android_doc}
\bibfield{author}{\bibinfo{person}{Google}.} \bibinfo{year}{2018}\natexlab{c}.
\newblock \bibinfo{title}{{Android Developers Documentation}}.
\newblock
\newblock
\urldef\tempurl%
\url{https://developer.android.com}
\showURL{%
Retrieved 2018-7 from \tempurl}


\bibitem[\protect\citeauthoryear{Google}{Google}{2018d}]%
        {lintchecks}
\bibfield{author}{\bibinfo{person}{Google}.} \bibinfo{year}{2018}\natexlab{d}.
\newblock \bibinfo{title}{{Android Lint Checks}}.
\newblock
\newblock
\urldef\tempurl%
\url{http://tools.android.com/tips/lint-checks}
\showURL{%
Retrieved 2018-7 from \tempurl}


\bibitem[\protect\citeauthoryear{Google}{Google}{2018e}]%
        {Android}
\bibfield{author}{\bibinfo{person}{Google}.} \bibinfo{year}{2018}\natexlab{e}.
\newblock \bibinfo{title}{{Android Platform}}.
\newblock
\newblock
\urldef\tempurl%
\url{https://www.android.com/}
\showURL{%
Retrieved 2018-7 from \tempurl}


\bibitem[\protect\citeauthoryear{Google}{Google}{2018f}]%
        {uiautomator}
\bibfield{author}{\bibinfo{person}{Google}.} \bibinfo{year}{2018}\natexlab{f}.
\newblock \bibinfo{title}{{Android UI Automator}}.
\newblock
\newblock
\urldef\tempurl%
\url{http://developer.android.com/tools/help/uiautomator/index.html}
\showURL{%
Retrieved 2018-7 from \tempurl}


\bibitem[\protect\citeauthoryear{Google}{Google}{2018g}]%
        {asynctask}
\bibfield{author}{\bibinfo{person}{Google}.} \bibinfo{year}{2018}\natexlab{g}.
\newblock \bibinfo{title}{{AsyncTask}}.
\newblock
\newblock
\urldef\tempurl%
\url{https://developer.android.com/reference/android/os/AsyncTask.html}
\showURL{%
Retrieved 2018-7 from \tempurl}


\bibitem[\protect\citeauthoryear{Google}{Google}{2018h}]%
        {dialog}
\bibfield{author}{\bibinfo{person}{Google}.} \bibinfo{year}{2018}\natexlab{h}.
\newblock \bibinfo{title}{{Dialogs}}.
\newblock
\newblock
\urldef\tempurl%
\url{https://developer.android.com/guide/topics/ui/dialogs.html}
\showURL{%
Retrieved 2018-7 from \tempurl}


\bibitem[\protect\citeauthoryear{Google}{Google}{2018i}]%
        {droidracer}
\bibfield{author}{\bibinfo{person}{Google}.} \bibinfo{year}{2018}\natexlab{i}.
\newblock \bibinfo{title}{{DroidRacer}}.
\newblock
\newblock
\urldef\tempurl%
\url{https://bitbucket.org/iiscseal/droidracer}
\showURL{%
Retrieved 2018-7 from \tempurl}


\bibitem[\protect\citeauthoryear{Google}{Google}{2018j}]%
        {fragment}
\bibfield{author}{\bibinfo{person}{Google}.} \bibinfo{year}{2018}\natexlab{j}.
\newblock \bibinfo{title}{{Fragment Lifecycle}}.
\newblock
\newblock
\urldef\tempurl%
\url{https://developer.android.com/guide/components/fragments.html}
\showURL{%
Retrieved 2018-7 from \tempurl}


\bibitem[\protect\citeauthoryear{Google}{Google}{2018k}]%
        {stateloss}
\bibfield{author}{\bibinfo{person}{Google}.} \bibinfo{year}{2018}\natexlab{k}.
\newblock \bibinfo{title}{{Fragment Transactions and Activity State Loss}}.
\newblock
\newblock
\urldef\tempurl%
\url{https://www.androiddesignpatterns.com/2013/08/fragment-transaction-commit-state-loss.html}
\showURL{%
Retrieved 2018-7 from \tempurl}


\bibitem[\protect\citeauthoryear{Google}{Google}{2018l}]%
        {handler}
\bibfield{author}{\bibinfo{person}{Google}.} \bibinfo{year}{2018}\natexlab{l}.
\newblock \bibinfo{title}{{Handler}}.
\newblock
\newblock
\urldef\tempurl%
\url{https://developer.android.com/reference/android/os/Handler.html}
\showURL{%
Retrieved 2018-7 from \tempurl}


\bibitem[\protect\citeauthoryear{Google}{Google}{2018m}]%
        {handlerthread}
\bibfield{author}{\bibinfo{person}{Google}.} \bibinfo{year}{2018}\natexlab{m}.
\newblock \bibinfo{title}{{HandlerThread}}.
\newblock
\newblock
\urldef\tempurl%
\url{https://developer.android.com/reference/android/os/HandlerThread.html}
\showURL{%
Retrieved 2018-7 from \tempurl}


\bibitem[\protect\citeauthoryear{Google}{Google}{2018n}]%
        {intentservice}
\bibfield{author}{\bibinfo{person}{Google}.} \bibinfo{year}{2018}\natexlab{n}.
\newblock \bibinfo{title}{{IntentService}}.
\newblock
\newblock
\urldef\tempurl%
\url{https://developer.android.com/reference/android/app/IntentService.html}
\showURL{%
Retrieved 2018-7 from \tempurl}


\bibitem[\protect\citeauthoryear{Google}{Google}{2018o}]%
        {lint}
\bibfield{author}{\bibinfo{person}{Google}.} \bibinfo{year}{2018}\natexlab{o}.
\newblock \bibinfo{title}{{Lint}}.
\newblock
\newblock
\urldef\tempurl%
\url{https://developer.android.com/studio/write/lint.html}
\showURL{%
Retrieved 2018-7 from \tempurl}


\bibitem[\protect\citeauthoryear{Google}{Google}{2018p}]%
        {listview}
\bibfield{author}{\bibinfo{person}{Google}.} \bibinfo{year}{2018}\natexlab{p}.
\newblock \bibinfo{title}{{List View}}.
\newblock
\newblock
\urldef\tempurl%
\url{https://developer.android.com/guide/topics/ui/layout/listview.html}
\showURL{%
Retrieved 2018-7 from \tempurl}


\bibitem[\protect\citeauthoryear{Google}{Google}{2018q}]%
        {loader}
\bibfield{author}{\bibinfo{person}{Google}.} \bibinfo{year}{2018}\natexlab{q}.
\newblock \bibinfo{title}{{Loaders}}.
\newblock
\newblock
\urldef\tempurl%
\url{https://developer.android.com/guide/components/loaders.html}
\showURL{%
Retrieved 2018-7 from \tempurl}


\bibitem[\protect\citeauthoryear{Google}{Google}{2018r}]%
        {looper}
\bibfield{author}{\bibinfo{person}{Google}.} \bibinfo{year}{2018}\natexlab{r}.
\newblock \bibinfo{title}{{Looper}}.
\newblock
\newblock
\urldef\tempurl%
\url{https://developer.android.com/reference/android/os/Looper.html}
\showURL{%
Retrieved 2018-7 from \tempurl}


\bibitem[\protect\citeauthoryear{Google}{Google}{2018s}]%
        {Monkey}
\bibfield{author}{\bibinfo{person}{Google}.} \bibinfo{year}{2018}\natexlab{s}.
\newblock \bibinfo{title}{{Monkey}}.
\newblock
\newblock
\urldef\tempurl%
\url{http://developer.android.com/tools/help/monkey.html}
\showURL{%
Retrieved 2018-7 from \tempurl}


\bibitem[\protect\citeauthoryear{Google}{Google}{2018t}]%
        {thread}
\bibfield{author}{\bibinfo{person}{Google}.} \bibinfo{year}{2018}\natexlab{t}.
\newblock \bibinfo{title}{{Processes and Threads}}.
\newblock
\newblock
\urldef\tempurl%
\url{https://developer.android.com/guide/components/processes-and-threads.html}
\showURL{%
Retrieved 2018-7 from \tempurl}


\bibitem[\protect\citeauthoryear{Google}{Google}{2018u}]%
        {thread1}
\bibfield{author}{\bibinfo{person}{Google}.} \bibinfo{year}{2018}\natexlab{u}.
\newblock \bibinfo{title}{{Thread}}.
\newblock
\newblock
\urldef\tempurl%
\url{https://developer.android.com/reference/java/lang/Thread.html}
\showURL{%
Retrieved 2018-7 from \tempurl}


\bibitem[\protect\citeauthoryear{Google}{Google}{2018v}]%
        {threadpool}
\bibfield{author}{\bibinfo{person}{Google}.} \bibinfo{year}{2018}\natexlab{v}.
\newblock \bibinfo{title}{{ThreadPoolExecutor}}.
\newblock
\newblock
\urldef\tempurl%
\url{https://developer.android.com/reference/java/util/concurrent/ThreadPoolExecutor.html}
\showURL{%
Retrieved 2018-7 from \tempurl}


\bibitem[\protect\citeauthoryear{Google}{Google}{2018w}]%
        {toast}
\bibfield{author}{\bibinfo{person}{Google}.} \bibinfo{year}{2018}\natexlab{w}.
\newblock \bibinfo{title}{{Toasts}}.
\newblock
\newblock
\urldef\tempurl%
\url{https://developer.android.com/guide/topics/ui/notifiers/toasts.html}
\showURL{%
Retrieved 2018-7 from \tempurl}


\bibitem[\protect\citeauthoryear{Graziussi}{Graziussi}{2016}]%
        {lifecycle16}
\bibfield{author}{\bibinfo{person}{Simone Graziussi}.}
  \bibinfo{year}{2016}\natexlab{}.
\newblock \emph{\bibinfo{title}{Lifecycle and Event-Based Testing for Android
  Applications}}.
\newblock \bibinfo{thesistype}{Master's\ thesis}. \bibinfo{school}{School Of
  Industrial Engineering and Information, Politecnico}.
\newblock


\bibitem[\protect\citeauthoryear{Hsiao, Pereira, Yu, Pokam, Narayanasamy, Chen,
  Kong, and Flinn}{Hsiao et~al\mbox{.}}{2014}]%
        {HsiaoPYPNCKF14}
\bibfield{author}{\bibinfo{person}{Chun{-}Hung Hsiao},
  \bibinfo{person}{Cristiano Pereira}, \bibinfo{person}{Jie Yu},
  \bibinfo{person}{Gilles Pokam}, \bibinfo{person}{Satish Narayanasamy},
  \bibinfo{person}{Peter~M. Chen}, \bibinfo{person}{Ziyun Kong}, {and}
  \bibinfo{person}{Jason Flinn}.} \bibinfo{year}{2014}\natexlab{}.
\newblock \showarticletitle{Race detection for event-driven mobile
  applications}. In \bibinfo{booktitle}{\emph{{ACM} {SIGPLAN} Conference on
  Programming Language Design and Implementation, {PLDI} '14, Edinburgh, United
  Kingdom - June 09 - 11, 2014}}. \bibinfo{pages}{326--336}.
\newblock


\bibitem[\protect\citeauthoryear{Hu and Neamtiu}{Hu and Neamtiu}{2011}]%
        {Hu11}
\bibfield{author}{\bibinfo{person}{Cuixiong Hu} {and} \bibinfo{person}{Iulian
  Neamtiu}.} \bibinfo{year}{2011}\natexlab{}.
\newblock \showarticletitle{Automating GUI Testing for Android Applications}.
  In \bibinfo{booktitle}{\emph{Proceedings of the 6th International Workshop on
  Automation of Software Test}} \emph{(\bibinfo{series}{AST '11})}.
  \bibinfo{publisher}{ACM}, \bibinfo{address}{New York, NY, USA},
  \bibinfo{pages}{77--83}.
\newblock
\showISBNx{978-1-4503-0592-1}


\bibitem[\protect\citeauthoryear{Hu, Neamtiu, and Alavi}{Hu
  et~al\mbox{.}}{2016}]%
        {Hu2016}
\bibfield{author}{\bibinfo{person}{Yongjian Hu}, \bibinfo{person}{Iulian
  Neamtiu}, {and} \bibinfo{person}{Arash Alavi}.}
  \bibinfo{year}{2016}\natexlab{}.
\newblock \showarticletitle{Automatically Verifying and Reproducing Event-based
  Races in Android Apps}. In \bibinfo{booktitle}{\emph{Proceedings of the 25th
  International Symposium on Software Testing and Analysis}}
  \emph{(\bibinfo{series}{ISSTA 2016})}. \bibinfo{publisher}{ACM},
  \bibinfo{address}{New York, NY, USA}, \bibinfo{pages}{377--388}.
\newblock
\showISBNx{978-1-4503-4390-9}


\bibitem[\protect\citeauthoryear{JDK}{JDK}{2018}]%
        {Swing}
\bibfield{author}{\bibinfo{person}{JDK}.} \bibinfo{year}{2018}\natexlab{}.
\newblock \bibinfo{title}{{JDK Swing Platform}}.
\newblock
\newblock
\urldef\tempurl%
\url{https://docs.oracle.com/javase/7/docs/technotes/guides/swing/}
\showURL{%
Retrieved 2018-7 from \tempurl}


\bibitem[\protect\citeauthoryear{Jensen, Prasad, and M{\o}ller}{Jensen
  et~al\mbox{.}}{2013}]%
        {JensenPM13}
\bibfield{author}{\bibinfo{person}{Casper~Svenning Jensen},
  \bibinfo{person}{Mukul~R. Prasad}, {and} \bibinfo{person}{Anders M{\o}ller}.}
  \bibinfo{year}{2013}\natexlab{}.
\newblock \showarticletitle{Automated testing with targeted event sequence
  generation}. In \bibinfo{booktitle}{\emph{International Symposium on Software
  Testing and Analysis, {ISSTA} '13, Lugano, Switzerland, July 15-20, 2013}}.
  \bibinfo{pages}{67--77}.
\newblock


\bibitem[\protect\citeauthoryear{Keng, Jiang, Wee, and Balan}{Keng
  et~al\mbox{.}}{2016}]%
        {Keng2016}
\bibfield{author}{\bibinfo{person}{Joseph Chan~Joo Keng},
  \bibinfo{person}{Lingxiao Jiang}, \bibinfo{person}{Tan~Kiat Wee}, {and}
  \bibinfo{person}{Rajesh~Krishna Balan}.} \bibinfo{year}{2016}\natexlab{}.
\newblock \showarticletitle{Graph-aided Directed Testing of Android
  Applications for Checking Runtime Privacy Behaviours}. In
  \bibinfo{booktitle}{\emph{Proceedings of the 11th International Workshop on
  Automation of Software Test}} \emph{(\bibinfo{series}{AST '16})}.
  \bibinfo{publisher}{ACM}, \bibinfo{address}{New York, NY, USA},
  \bibinfo{pages}{57--63}.
\newblock
\showISBNx{978-1-4503-4151-6}


\bibitem[\protect\citeauthoryear{Kochhar, Thung, Nagappan, Zimmermann, and
  Lo}{Kochhar et~al\mbox{.}}{2015}]%
        {KochharTNZL15}
\bibfield{author}{\bibinfo{person}{Pavneet~Singh Kochhar},
  \bibinfo{person}{Ferdian Thung}, \bibinfo{person}{Nachiappan Nagappan},
  \bibinfo{person}{Thomas Zimmermann}, {and} \bibinfo{person}{David Lo}.}
  \bibinfo{year}{2015}\natexlab{}.
\newblock \showarticletitle{Understanding the Test Automation Culture of App
  Developers}. In \bibinfo{booktitle}{\emph{8th {IEEE} International Conference
  on Software Testing, Verification and Validation, {ICST} 2015, Graz, Austria,
  April 13-17, 2015}}. \bibinfo{pages}{1--10}.
\newblock


\bibitem[\protect\citeauthoryear{Li, Jiang, Gu, Xu, Ma, Ma, and Lu}{Li
  et~al\mbox{.}}{2016}]%
        {Li0GXMML16}
\bibfield{author}{\bibinfo{person}{Qiwei Li}, \bibinfo{person}{Yanyan Jiang},
  \bibinfo{person}{Tianxiao Gu}, \bibinfo{person}{Chang Xu},
  \bibinfo{person}{Jun Ma}, \bibinfo{person}{Xiaoxing Ma}, {and}
  \bibinfo{person}{Jian Lu}.} \bibinfo{year}{2016}\natexlab{}.
\newblock \showarticletitle{Effectively Manifesting Concurrency Bugs in Android
  Apps}. In \bibinfo{booktitle}{\emph{23rd Asia-Pacific Software Engineering
  Conference, {APSEC} 2016, Hamilton, New Zealand, December 6-9, 2016}}.
  \bibinfo{pages}{209--216}.
\newblock


\bibitem[\protect\citeauthoryear{Lin, Okur, and Dig}{Lin et~al\mbox{.}}{2015}]%
        {LinOD15}
\bibfield{author}{\bibinfo{person}{Yu Lin}, \bibinfo{person}{Semih Okur}, {and}
  \bibinfo{person}{Danny Dig}.} \bibinfo{year}{2015}\natexlab{}.
\newblock \showarticletitle{Study and Refactoring of Android Asynchronous
  Programming {(T)}}. In \bibinfo{booktitle}{\emph{30th {IEEE/ACM}
  International Conference on Automated Software Engineering, {ASE} 2015,
  Lincoln, NE, USA, November 9-13, 2015}}. \bibinfo{pages}{224--235}.
\newblock


\bibitem[\protect\citeauthoryear{Lin, Radoi, and Dig}{Lin
  et~al\mbox{.}}{2014}]%
        {Lin14}
\bibfield{author}{\bibinfo{person}{Yu Lin}, \bibinfo{person}{Cosmin Radoi},
  {and} \bibinfo{person}{Danny Dig}.} \bibinfo{year}{2014}\natexlab{}.
\newblock \showarticletitle{Retrofitting Concurrency for Android Applications
  Through Refactoring}. In \bibinfo{booktitle}{\emph{Proceedings of the 22Nd
  ACM SIGSOFT International Symposium on Foundations of Software Engineering}}
  \emph{(\bibinfo{series}{FSE 2014})}. \bibinfo{publisher}{ACM},
  \bibinfo{address}{New York, NY, USA}, \bibinfo{pages}{341--352}.
\newblock
\showISBNx{978-1-4503-3056-5}


\bibitem[\protect\citeauthoryear{Linares-V\'{a}squez, Bavota, Tufano, Moran,
  Di~Penta, Vendome, Bernal-C\'{a}rdenas, and Poshyvanyk}{Linares-V\'{a}squez
  et~al\mbox{.}}{2017}]%
        {Linares2017}
\bibfield{author}{\bibinfo{person}{Mario Linares-V\'{a}squez},
  \bibinfo{person}{Gabriele Bavota}, \bibinfo{person}{Michele Tufano},
  \bibinfo{person}{Kevin Moran}, \bibinfo{person}{Massimiliano Di~Penta},
  \bibinfo{person}{Christopher Vendome}, \bibinfo{person}{Carlos
  Bernal-C\'{a}rdenas}, {and} \bibinfo{person}{Denys Poshyvanyk}.}
  \bibinfo{year}{2017}\natexlab{}.
\newblock \showarticletitle{Enabling Mutation Testing for Android Apps}. In
  \bibinfo{booktitle}{\emph{Proceedings of the 2017 11th Joint Meeting on
  Foundations of Software Engineering}} \emph{(\bibinfo{series}{ESEC/FSE
  2017})}. \bibinfo{publisher}{ACM}, \bibinfo{address}{New York, NY, USA},
  \bibinfo{pages}{233--244}.
\newblock
\showISBNx{978-1-4503-5105-8}


\bibitem[\protect\citeauthoryear{Ma, Khoo, Foster, and Hicks}{Ma
  et~al\mbox{.}}{2011}]%
        {MaKFH11}
\bibfield{author}{\bibinfo{person}{Kin{-}Keung Ma}, \bibinfo{person}{Yit~Phang
  Khoo}, \bibinfo{person}{Jeffrey~S. Foster}, {and} \bibinfo{person}{Michael
  Hicks}.} \bibinfo{year}{2011}\natexlab{}.
\newblock \showarticletitle{Directed Symbolic Execution}. In
  \bibinfo{booktitle}{\emph{Static Analysis - 18th International Symposium,
  {SAS} 2011, Venice, Italy, September 14-16, 2011. Proceedings}}.
  \bibinfo{pages}{95--111}.
\newblock


\bibitem[\protect\citeauthoryear{Machiry, Tahiliani, and Naik}{Machiry
  et~al\mbox{.}}{2013}]%
        {MachiryTN13}
\bibfield{author}{\bibinfo{person}{Aravind Machiry}, \bibinfo{person}{Rohan
  Tahiliani}, {and} \bibinfo{person}{Mayur Naik}.}
  \bibinfo{year}{2013}\natexlab{}.
\newblock \showarticletitle{Dynodroid: an input generation system for {Android}
  apps}. In \bibinfo{booktitle}{\emph{Joint Meeting of the European Software
  Engineering Conference and the {ACM} {SIGSOFT} Symposium on the Foundations
  of Software Engineering, ESEC/FSE'13, Saint Petersburg, Russian Federation,
  August 18-26, 2013}}. \bibinfo{pages}{224--234}.
\newblock


\bibitem[\protect\citeauthoryear{Mahmood, Mirzaei, and Malek}{Mahmood
  et~al\mbox{.}}{2014}]%
        {MahmoodMM14}
\bibfield{author}{\bibinfo{person}{Riyadh Mahmood}, \bibinfo{person}{Nariman
  Mirzaei}, {and} \bibinfo{person}{Sam Malek}.}
  \bibinfo{year}{2014}\natexlab{}.
\newblock \showarticletitle{{EvoDroid}: segmented evolutionary testing of
  {Android} apps}. In \bibinfo{booktitle}{\emph{Proceedings of the 22nd {ACM}
  {SIGSOFT} International Symposium on Foundations of Software Engineering,
  (FSE-22), Hong Kong, China, November 16 - 22, 2014}}.
  \bibinfo{pages}{599--609}.
\newblock


\bibitem[\protect\citeauthoryear{Maiya, Kanade, and Majumdar}{Maiya
  et~al\mbox{.}}{2014}]%
        {MaiyaKM14}
\bibfield{author}{\bibinfo{person}{Pallavi Maiya}, \bibinfo{person}{Aditya
  Kanade}, {and} \bibinfo{person}{Rupak Majumdar}.}
  \bibinfo{year}{2014}\natexlab{}.
\newblock \showarticletitle{Race detection for Android applications}. In
  \bibinfo{booktitle}{\emph{{ACM} {SIGPLAN} Conference on Programming Language
  Design and Implementation, {PLDI} '14, Edinburgh, United Kingdom - June 09 -
  11, 2014}}. \bibinfo{pages}{316--325}.
\newblock


\bibitem[\protect\citeauthoryear{Mao, Harman, and Jia}{Mao
  et~al\mbox{.}}{2016}]%
        {MaoHJ16}
\bibfield{author}{\bibinfo{person}{Ke Mao}, \bibinfo{person}{Mark Harman},
  {and} \bibinfo{person}{Yue Jia}.} \bibinfo{year}{2016}\natexlab{}.
\newblock \showarticletitle{Sapienz: multi-objective automated testing for
  Android applications}. In \bibinfo{booktitle}{\emph{Proceedings of the 25th
  International Symposium on Software Testing and Analysis, {ISSTA} 2016,
  Saarbr{\"{u}}cken, Germany, July 18-20, 2016}}. \bibinfo{pages}{94--105}.
\newblock


\bibitem[\protect\citeauthoryear{Mao, Harman, and Jia}{Mao
  et~al\mbox{.}}{2017}]%
        {MaoHJ17}
\bibfield{author}{\bibinfo{person}{Ke Mao}, \bibinfo{person}{Mark Harman},
  {and} \bibinfo{person}{Yue Jia}.} \bibinfo{year}{2017}\natexlab{}.
\newblock \showarticletitle{Crowd intelligence enhances automated mobile
  testing}. In \bibinfo{booktitle}{\emph{Proceedings of the 32nd {IEEE/ACM}
  International Conference on Automated Software Engineering, {ASE} 2017,
  Urbana, IL, USA, October 30 - November 03, 2017}}. \bibinfo{pages}{16--26}.
\newblock


\bibitem[\protect\citeauthoryear{Mirzaei, Garcia, Bagheri, Sadeghi, and
  Malek}{Mirzaei et~al\mbox{.}}{2016}]%
        {Mirzaei16}
\bibfield{author}{\bibinfo{person}{Nariman Mirzaei}, \bibinfo{person}{Joshua
  Garcia}, \bibinfo{person}{Hamid Bagheri}, \bibinfo{person}{Alireza Sadeghi},
  {and} \bibinfo{person}{Sam Malek}.} \bibinfo{year}{2016}\natexlab{}.
\newblock \showarticletitle{Reducing Combinatorics in GUI Testing of Android
  Applications}. In \bibinfo{booktitle}{\emph{Proceedings of the 38th
  International Conference on Software Engineering}}
  \emph{(\bibinfo{series}{ICSE '16})}. \bibinfo{publisher}{ACM},
  \bibinfo{address}{New York, NY, USA}, \bibinfo{pages}{559--570}.
\newblock


\bibitem[\protect\citeauthoryear{Mirzaei, Malek, Pasareanu, Esfahani, and
  Mahmood}{Mirzaei et~al\mbox{.}}{2012}]%
        {MirzaeiMPEM12}
\bibfield{author}{\bibinfo{person}{Nariman Mirzaei}, \bibinfo{person}{Sam
  Malek}, \bibinfo{person}{Corina~S. Pasareanu}, \bibinfo{person}{Naeem
  Esfahani}, {and} \bibinfo{person}{Riyadh Mahmood}.}
  \bibinfo{year}{2012}\natexlab{}.
\newblock \showarticletitle{Testing {Android} apps through symbolic execution}.
\newblock \bibinfo{journal}{\emph{{ACM} {SIGSOFT} Software Engineering Notes}}
  \bibinfo{volume}{37}, \bibinfo{number}{6} (\bibinfo{year}{2012}),
  \bibinfo{pages}{1--5}.
\newblock


\bibitem[\protect\citeauthoryear{Moran, V{\'{a}}squez, Bernal{-}C{\'{a}}rdenas,
  Vendome, and Poshyvanyk}{Moran et~al\mbox{.}}{2016}]%
        {MoranVBVP16}
\bibfield{author}{\bibinfo{person}{Kevin Moran}, \bibinfo{person}{Mario~Linares
  V{\'{a}}squez}, \bibinfo{person}{Carlos Bernal{-}C{\'{a}}rdenas},
  \bibinfo{person}{Christopher Vendome}, {and} \bibinfo{person}{Denys
  Poshyvanyk}.} \bibinfo{year}{2016}\natexlab{}.
\newblock \showarticletitle{Automatically Discovering, Reporting and
  Reproducing Android Application Crashes}. In \bibinfo{booktitle}{\emph{2016
  {IEEE} International Conference on Software Testing, Verification and
  Validation, {ICST} 2016, Chicago, IL, USA, April 11-15, 2016}}.
  \bibinfo{pages}{33--44}.
\newblock


\bibitem[\protect\citeauthoryear{Octeau, Luchaup, Dering, Jha, and
  McDaniel}{Octeau et~al\mbox{.}}{2015}]%
        {Octeau15}
\bibfield{author}{\bibinfo{person}{Damien Octeau}, \bibinfo{person}{Daniel
  Luchaup}, \bibinfo{person}{Matthew Dering}, \bibinfo{person}{Somesh Jha},
  {and} \bibinfo{person}{Patrick McDaniel}.} \bibinfo{year}{2015}\natexlab{}.
\newblock \showarticletitle{Composite Constant Propagation: Application to
  Android Inter-component Communication Analysis}. In
  \bibinfo{booktitle}{\emph{Proceedings of the 37th International Conference on
  Software Engineering - Volume 1}} \emph{(\bibinfo{series}{ICSE '15})}.
  \bibinfo{publisher}{IEEE Press}, \bibinfo{address}{Piscataway, NJ, USA},
  \bibinfo{pages}{77--88}.
\newblock
\showISBNx{978-1-4799-1934-5}


\bibitem[\protect\citeauthoryear{Ozkan, Emmi, and Tasiran}{Ozkan
  et~al\mbox{.}}{2015}]%
        {OzkanET15}
\bibfield{author}{\bibinfo{person}{Burcu~Kulahcioglu Ozkan},
  \bibinfo{person}{Michael Emmi}, {and} \bibinfo{person}{Serdar Tasiran}.}
  \bibinfo{year}{2015}\natexlab{}.
\newblock \showarticletitle{Systematic Asynchrony Bug Exploration for Android
  Apps}. In \bibinfo{booktitle}{\emph{Computer Aided Verification - 27th
  International Conference, {CAV} 2015, San Francisco, CA, USA, July 18-24,
  2015, Proceedings, Part {I}}}. \bibinfo{pages}{455--461}.
\newblock


\bibitem[\protect\citeauthoryear{PMD}{PMD}{2018}]%
        {pmd}
\bibfield{author}{\bibinfo{person}{PMD}.} \bibinfo{year}{2018}\natexlab{}.
\newblock \bibinfo{title}{{PMD}}.
\newblock
\newblock
\urldef\tempurl%
\url{https://pmd.github.io/}
\showURL{%
Retrieved 2018-7 from \tempurl}


\bibitem[\protect\citeauthoryear{Qt}{Qt}{2018}]%
        {Qt}
\bibfield{author}{\bibinfo{person}{Qt}.} \bibinfo{year}{2018}\natexlab{}.
\newblock \bibinfo{title}{{Qt}}.
\newblock
\newblock
\urldef\tempurl%
\url{https://www.qt.io/}
\showURL{%
Retrieved 2018-7 from \tempurl}


\bibitem[\protect\citeauthoryear{Rasthofer, Arzt, Triller, and
  Pradel}{Rasthofer et~al\mbox{.}}{2017}]%
        {Rasthofer2017}
\bibfield{author}{\bibinfo{person}{Siegfried Rasthofer},
  \bibinfo{person}{Steven Arzt}, \bibinfo{person}{Stefan Triller}, {and}
  \bibinfo{person}{Michael Pradel}.} \bibinfo{year}{2017}\natexlab{}.
\newblock \showarticletitle{Making Malory Behave Maliciously: Targeted Fuzzing
  of Android Execution Environments}. In \bibinfo{booktitle}{\emph{Proceedings
  of the 39th International Conference on Software Engineering}}
  \emph{(\bibinfo{series}{ICSE '17})}. \bibinfo{pages}{300--311}.
\newblock
\showISBNx{978-1-5386-3868-2}


\bibitem[\protect\citeauthoryear{Rountev and Yan}{Rountev and Yan}{2014}]%
        {RountevY14}
\bibfield{author}{\bibinfo{person}{Atanas Rountev} {and}
  \bibinfo{person}{Dacong Yan}.} \bibinfo{year}{2014}\natexlab{}.
\newblock \showarticletitle{Static Reference Analysis for {GUI} Objects in
  Android Software}. In \bibinfo{booktitle}{\emph{12th Annual {IEEE/ACM}
  International Symposium on Code Generation and Optimization, {CGO} 2014,
  Orlando, FL, USA, February 15-19, 2014}}. \bibinfo{pages}{143}.
\newblock


\bibitem[\protect\citeauthoryear{Safi, Shahbazian, Halfond, and
  Medvidovic}{Safi et~al\mbox{.}}{2015}]%
        {Safi15}
\bibfield{author}{\bibinfo{person}{Gholamreza Safi}, \bibinfo{person}{Arman
  Shahbazian}, \bibinfo{person}{William G.~J. Halfond}, {and}
  \bibinfo{person}{Nenad Medvidovic}.} \bibinfo{year}{2015}\natexlab{}.
\newblock \showarticletitle{Detecting Event Anomalies in Event-based Systems}.
  In \bibinfo{booktitle}{\emph{Proceedings of the 2015 10th Joint Meeting on
  Foundations of Software Engineering}} \emph{(\bibinfo{series}{ESEC/FSE
  2015})}. \bibinfo{publisher}{ACM}, \bibinfo{address}{New York, NY, USA},
  \bibinfo{pages}{25--37}.
\newblock
\showISBNx{978-1-4503-3675-8}


\bibitem[\protect\citeauthoryear{Santhiar, Kaleeswaran, and Kanade}{Santhiar
  et~al\mbox{.}}{2016}]%
        {SanthiarKK16}
\bibfield{author}{\bibinfo{person}{Anirudh Santhiar}, \bibinfo{person}{Shalini
  Kaleeswaran}, {and} \bibinfo{person}{Aditya Kanade}.}
  \bibinfo{year}{2016}\natexlab{}.
\newblock \showarticletitle{Efficient race detection in the presence of
  programmatic event loops}. In \bibinfo{booktitle}{\emph{Proceedings of the
  25th International Symposium on Software Testing and Analysis, {ISSTA} 2016,
  Saarbr{\"{u}}cken, Germany, July 18-20, 2016}}. \bibinfo{pages}{366--376}.
\newblock


\bibitem[\protect\citeauthoryear{Sch{\"{u}}tte, Fedler, and
  Titze}{Sch{\"{u}}tte et~al\mbox{.}}{2015}]%
        {SchutteFT15}
\bibfield{author}{\bibinfo{person}{Julian Sch{\"{u}}tte},
  \bibinfo{person}{Rafael Fedler}, {and} \bibinfo{person}{Dennis Titze}.}
  \bibinfo{year}{2015}\natexlab{}.
\newblock \showarticletitle{ConDroid: Targeted Dynamic Analysis of Android
  Applications}. In \bibinfo{booktitle}{\emph{29th {IEEE} International
  Conference on Advanced Information Networking and Applications, {AINA} 2015,
  Gwangju, South Korea, March 24-27, 2015}}. \bibinfo{pages}{571--578}.
\newblock


\bibitem[\protect\citeauthoryear{Shan, Azim, and Neamtiu}{Shan
  et~al\mbox{.}}{2016}]%
        {Shan16}
\bibfield{author}{\bibinfo{person}{Zhiyong Shan}, \bibinfo{person}{Tanzirul
  Azim}, {and} \bibinfo{person}{Iulian Neamtiu}.}
  \bibinfo{year}{2016}\natexlab{}.
\newblock \showarticletitle{Finding Resume and Restart Errors in Android
  Applications}. In \bibinfo{booktitle}{\emph{Proceedings of the 2016 ACM
  SIGPLAN International Conference on Object-Oriented Programming, Systems,
  Languages, and Applications}} \emph{(\bibinfo{series}{OOPSLA 2016})}.
  \bibinfo{publisher}{ACM}, \bibinfo{address}{New York, NY, USA},
  \bibinfo{pages}{864--880}.
\newblock
\showISBNx{978-1-4503-4444-9}


\bibitem[\protect\citeauthoryear{Song, Qian, and Huang}{Song
  et~al\mbox{.}}{2017}]%
        {Song2017}
\bibfield{author}{\bibinfo{person}{Wei Song}, \bibinfo{person}{Xiangxing Qian},
  {and} \bibinfo{person}{Jeff Huang}.} \bibinfo{year}{2017}\natexlab{}.
\newblock \showarticletitle{EHBDroid: Beyond GUI Testing for Android
  Applications}. In \bibinfo{booktitle}{\emph{Proceedings of the 32Nd IEEE/ACM
  International Conference on Automated Software Engineering}}
  \emph{(\bibinfo{series}{ASE 2017})}. \bibinfo{publisher}{IEEE Press},
  \bibinfo{address}{Piscataway, NJ, USA}, \bibinfo{pages}{27--37}.
\newblock
\showISBNx{978-1-5386-2684-9}


\bibitem[\protect\citeauthoryear{Su}{Su}{2016}]%
        {FSMdroid16}
\bibfield{author}{\bibinfo{person}{Ting Su}.} \bibinfo{year}{2016}\natexlab{}.
\newblock \showarticletitle{FSMdroid: guided {GUI} testing of android apps}. In
  \bibinfo{booktitle}{\emph{Proceedings of the 38th International Conference on
  Software Engineering, {ICSE} 2016, Austin, TX, USA, May 14-22, 2016 -
  Companion Volume}}. \bibinfo{pages}{689--691}.
\newblock


\bibitem[\protect\citeauthoryear{Su, Fu, Pu, He, and Su}{Su
  et~al\mbox{.}}{2015}]%
        {SuFPHS15}
\bibfield{author}{\bibinfo{person}{Ting Su}, \bibinfo{person}{Zhoulai Fu},
  \bibinfo{person}{Geguang Pu}, \bibinfo{person}{Jifeng He}, {and}
  \bibinfo{person}{Zhendong Su}.} \bibinfo{year}{2015}\natexlab{}.
\newblock \showarticletitle{Combining Symbolic Execution and Model Checking for
  Data Flow Testing}. In \bibinfo{booktitle}{\emph{37th {IEEE/ACM}
  International Conference on Software Engineering, {ICSE} 2015, Florence,
  Italy, May 16-24, 2015, Volume 1}}. \bibinfo{pages}{654--665}.
\newblock


\bibitem[\protect\citeauthoryear{Su, Meng, Chen, Wu, Yang, Yao, Pu, Liu, and
  Su}{Su et~al\mbox{.}}{2017}]%
        {Stoat2017}
\bibfield{author}{\bibinfo{person}{Ting Su}, \bibinfo{person}{Guozhu Meng},
  \bibinfo{person}{Yuting Chen}, \bibinfo{person}{Ke Wu},
  \bibinfo{person}{Weiming Yang}, \bibinfo{person}{Yao Yao},
  \bibinfo{person}{Geguang Pu}, \bibinfo{person}{Yang Liu}, {and}
  \bibinfo{person}{Zhendong Su}.} \bibinfo{year}{2017}\natexlab{}.
\newblock \showarticletitle{Guided, Stochastic Model-based GUI Testing of
  Android Apps}. In \bibinfo{booktitle}{\emph{Proceedings of the 2017 11th
  Joint Meeting on Foundations of Software Engineering}}
  \emph{(\bibinfo{series}{ESEC/FSE 2017})}. \bibinfo{publisher}{ACM},
  \bibinfo{address}{New York, NY, USA}, \bibinfo{pages}{245--256}.
\newblock
\showISBNx{978-1-4503-5105-8}


\bibitem[\protect\citeauthoryear{SWT}{SWT}{2018}]%
        {swt}
\bibfield{author}{\bibinfo{person}{SWT}.} \bibinfo{year}{2018}\natexlab{}.
\newblock \bibinfo{title}{{SWT}}.
\newblock
\newblock
\urldef\tempurl%
\url{https://www.eclipse.org/swt/}
\showURL{%
Retrieved 2018-7 from \tempurl}


\bibitem[\protect\citeauthoryear{Tang, Wu, Wei, and Zhong}{Tang
  et~al\mbox{.}}{2016}]%
        {Tang2016}
\bibfield{author}{\bibinfo{person}{Hongyin Tang}, \bibinfo{person}{Guoquan Wu},
  \bibinfo{person}{Jun Wei}, {and} \bibinfo{person}{Hua Zhong}.}
  \bibinfo{year}{2016}\natexlab{}.
\newblock \showarticletitle{Generating Test Cases to Expose Concurrency Bugs in
  Android Applications}. In \bibinfo{booktitle}{\emph{Proceedings of the 31st
  IEEE/ACM International Conference on Automated Software Engineering}}
  \emph{(\bibinfo{series}{ASE 2016})}. \bibinfo{publisher}{ACM},
  \bibinfo{address}{New York, NY, USA}, \bibinfo{pages}{648--653}.
\newblock
\showISBNx{978-1-4503-3845-5}


\bibitem[\protect\citeauthoryear{Vall{\'e}e-Rai, Co, Gagnon, Hendren, Lam, and
  Sundaresan}{Vall{\'e}e-Rai et~al\mbox{.}}{1999}]%
        {soot}
\bibfield{author}{\bibinfo{person}{Raja Vall{\'e}e-Rai}, \bibinfo{person}{Phong
  Co}, \bibinfo{person}{Etienne Gagnon}, \bibinfo{person}{Laurie Hendren},
  \bibinfo{person}{Patrick Lam}, {and} \bibinfo{person}{Vijay Sundaresan}.}
  \bibinfo{year}{1999}\natexlab{}.
\newblock \showarticletitle{Soot-a Java bytecode optimization framework}. In
  \bibinfo{booktitle}{\emph{Proceedings of the 1999 conference of the Centre
  for Advanced Studies on Collaborative research}}. IBM Press,
  \bibinfo{pages}{13}.
\newblock


\bibitem[\protect\citeauthoryear{van~der Merwe, van~der Merwe, and
  Visser}{van~der Merwe et~al\mbox{.}}{2012}]%
        {vanderMerwe2012}
\bibfield{author}{\bibinfo{person}{Heila van~der Merwe}, \bibinfo{person}{Brink
  van~der Merwe}, {and} \bibinfo{person}{Willem Visser}.}
  \bibinfo{year}{2012}\natexlab{}.
\newblock \showarticletitle{Verifying Android Applications Using Java
  PathFinder}.
\newblock \bibinfo{journal}{\emph{SIGSOFT Softw. Eng. Notes}}
  \bibinfo{volume}{37}, \bibinfo{number}{6} (\bibinfo{date}{Nov.}
  \bibinfo{year}{2012}), \bibinfo{pages}{1--5}.
\newblock
\showISSN{0163-5948}


\bibitem[\protect\citeauthoryear{V{\'{a}}squez, White, Bernal{-}C{\'{a}}rdenas,
  Moran, and Poshyvanyk}{V{\'{a}}squez et~al\mbox{.}}{2015}]%
        {VasquezWBMP15}
\bibfield{author}{\bibinfo{person}{Mario~Linares V{\'{a}}squez},
  \bibinfo{person}{Martin White}, \bibinfo{person}{Carlos
  Bernal{-}C{\'{a}}rdenas}, \bibinfo{person}{Kevin Moran}, {and}
  \bibinfo{person}{Denys Poshyvanyk}.} \bibinfo{year}{2015}\natexlab{}.
\newblock \showarticletitle{Mining Android App Usages for Generating Actionable
  GUI-Based Execution Scenarios}. In \bibinfo{booktitle}{\emph{12th {IEEE/ACM}
  Working Conference on Mining Software Repositories, {MSR} 2015, Florence,
  Italy, May 16-17, 2015}}. \bibinfo{pages}{111--122}.
\newblock


\bibitem[\protect\citeauthoryear{wxErlang}{wxErlang}{2018}]%
        {wxerlang}
\bibfield{author}{\bibinfo{person}{wxErlang}.} \bibinfo{year}{2018}\natexlab{}.
\newblock \bibinfo{title}{{wxErlang}}.
\newblock
\newblock
\urldef\tempurl%
\url{http://erlang.org/doc/apps/wx/index.html}
\showURL{%
Retrieved 2018-7 from \tempurl}


\bibitem[\protect\citeauthoryear{Yang, Prasad, and Xie}{Yang
  et~al\mbox{.}}{2013}]%
        {YangPX13}
\bibfield{author}{\bibinfo{person}{Wei Yang}, \bibinfo{person}{Mukul~R.
  Prasad}, {and} \bibinfo{person}{Tao Xie}.} \bibinfo{year}{2013}\natexlab{}.
\newblock \showarticletitle{A Grey-Box Approach for Automated GUI-Model
  Generation of Mobile Applications}. In \bibinfo{booktitle}{\emph{Fundamental
  Approaches to Software Engineering - 16th International Conference, {FASE}
  2013, Held as Part of the European Joint Conferences on Theory and Practice
  of Software, {ETAPS} 2013, Rome, Italy, March 16-24, 2013. Proceedings}}.
  \bibinfo{pages}{250--265}.
\newblock


\bibitem[\protect\citeauthoryear{Zaeem, Prasad, and Khurshid}{Zaeem
  et~al\mbox{.}}{2014}]%
        {Zaeem14}
\bibfield{author}{\bibinfo{person}{Razieh~Nokhbeh Zaeem},
  \bibinfo{person}{Mukul~R. Prasad}, {and} \bibinfo{person}{Sarfraz Khurshid}.}
  \bibinfo{year}{2014}\natexlab{}.
\newblock \showarticletitle{Automated Generation of Oracles for Testing
  User-Interaction Features of Mobile Apps}. In
  \bibinfo{booktitle}{\emph{Proceedings of the 2014 IEEE International
  Conference on Software Testing, Verification, and Validation}}
  \emph{(\bibinfo{series}{ICST '14})}. \bibinfo{publisher}{IEEE Computer
  Society}, \bibinfo{address}{Washington, DC, USA}, \bibinfo{pages}{183--192}.
\newblock
\showISBNx{978-1-4799-2255-0}


\bibitem[\protect\citeauthoryear{Zhang and Elbaum}{Zhang and Elbaum}{2014}]%
        {ZhangE14}
\bibfield{author}{\bibinfo{person}{Pingyu Zhang} {and}
  \bibinfo{person}{Sebastian~G. Elbaum}.} \bibinfo{year}{2014}\natexlab{}.
\newblock \showarticletitle{Amplifying Tests to Validate Exception Handling
  Code: An Extended Study in the Mobile Application Domain}.
\newblock \bibinfo{journal}{\emph{{ACM} Trans. Softw. Eng. Methodol.}}
  \bibinfo{volume}{23}, \bibinfo{number}{4} (\bibinfo{year}{2014}),
  \bibinfo{pages}{32:1--32:28}.
\newblock


\bibitem[\protect\citeauthoryear{Zhang, L\"{u}, and Ernst}{Zhang
  et~al\mbox{.}}{2012}]%
        {Zhang2012}
\bibfield{author}{\bibinfo{person}{Sai Zhang}, \bibinfo{person}{Hao L\"{u}},
  {and} \bibinfo{person}{Michael~D. Ernst}.} \bibinfo{year}{2012}\natexlab{}.
\newblock \showarticletitle{Finding Errors in Multithreaded GUI Applications}.
  In \bibinfo{booktitle}{\emph{Proceedings of the 2012 International Symposium
  on Software Testing and Analysis}} \emph{(\bibinfo{series}{ISSTA 2012})}.
  \bibinfo{pages}{243--253}.
\newblock
\showISBNx{978-1-4503-1454-1}


\bibitem[\protect\citeauthoryear{Zheng, Zhu, Dai, Gu, Gong, Han, and Zou}{Zheng
  et~al\mbox{.}}{2012}]%
        {Zheng12}
\bibfield{author}{\bibinfo{person}{Cong Zheng}, \bibinfo{person}{Shixiong Zhu},
  \bibinfo{person}{Shuaifu Dai}, \bibinfo{person}{Guofei Gu},
  \bibinfo{person}{Xiaorui Gong}, \bibinfo{person}{Xinhui Han}, {and}
  \bibinfo{person}{Wei Zou}.} \bibinfo{year}{2012}\natexlab{}.
\newblock \showarticletitle{SmartDroid: An Automatic System for Revealing
  UI-based Trigger Conditions in Android Applications}. In
  \bibinfo{booktitle}{\emph{Proceedings of the Second ACM Workshop on Security
  and Privacy in Smartphones and Mobile Devices}} \emph{(\bibinfo{series}{SPSM
  '12})}. \bibinfo{publisher}{ACM}, \bibinfo{address}{New York, NY, USA},
  \bibinfo{pages}{93--104}.
\newblock
\showISBNx{978-1-4503-1666-8}


\end{thebibliography}
